\documentclass[11pt,a4paper]{article}
\usepackage{jheppub}
\pdfoutput=1

\usepackage{epsfig,amsfonts,amstext,afterpage,psfrag,cancel}

\usepackage{slashed}
\usepackage{textcomp}
\usepackage{overpic}
\usepackage{booktabs,multirow}

\usepackage[table]{xcolor}

\newcommand{\crowcolor}{\rowcolor[rgb]{0.9,0.9,0.9}}
\newcommand{\prob}[2]{\text{prob}( #1 | #2 )}
\newcommand{\ewXL}{$ew\chi\mathcal{L}$}
\newcommand{\SM}{\mathrm{SM}}

\long\def\symbolfootnote[#1]#2{\begingroup%
\def\thefootnote{\fnsymbol{footnote}}\footnote[#1]{#2}\endgroup} 

\definecolor{color1}{RGB}{119,255,85}
\definecolor{color2}{RGB}{221,102,0}
\definecolor{color3}{RGB}{0,204,255}
\definecolor{color4}{RGB}{170,51,136}
\definecolor{color5}{RGB}{204,204,170}

\catcode`@=11
\def\@citex[#1]#2{%
  \if@filesw\immediate\write\@auxout{\string\citation{#2}}\fi
  \def\@citea{}\@cite{\@for\@citeb:=#2\do
    {\@citea\def\@citea{,\penalty\@m}\@ifundefined
      {b@\@citeb}{{\bf ?}\@warning
{Citation `\@citeb' on page \thepage \space undefined}}%
      \hbox{\csname b@\@citeb\endcsname}}}{#1}}
\def\citer{\@ifnextchar [{\@tempswatrue\@citexr}{\@tempswafalse\@citexr[]}}
  \def\@citexr[#1]#2{%
    \if@filesw\immediate\write\@auxout{\string\citation{#2}}\fi
    \def\@citea{}\@cite{\@for\@citeb:=#2\do
      {\@citea\def\@citea{--\penalty\@m}\@ifundefined
{b@\@citeb}{{\bf ?}\@warning
{Citation `\@citeb' on page \thepage \space undefined}}%
\hbox{\csname b@\@citeb\endcsname}}}{#1}}

\title{Current and future constraints on Higgs couplings in the nonlinear Effective Theory}

\author[a,b]{Jorge de Blas,}
\author[c]{Otto Eberhardt,}
\author[c,d]{and Claudius Krause}
 
\affiliation[a]{Dipartimento di Fisica e Astronomia ``Galileo Galilei'', Universit\`a di Padova, Via Marzolo 8, I-35131 Padova, Italy}
\affiliation[b]{INFN, Sezione di Padova, Via Marzolo 8, I-35131 Padova, Italy}
\affiliation[c]{IFIC, Universitat de Valencia-CSIC,\\ Apt. Correus 22085, E-46071 Valencia, Spain}
\affiliation[d]{Theoretical Physics Department, Fermi National Accelerator Laboratory, Batavia, IL, 60510, USA}

\emailAdd{Jorge.DeBlasMateo@pd.infn.it}
\emailAdd{Otto.Eberhardt@ific.uv.es}
\emailAdd{ckrause@fnal.gov}

\abstract{
We perform a Bayesian statistical analysis of the constraints on the nonlinear Effective Theory given by the Higgs electroweak chiral Lagrangian. 
We obtain bounds on the effective coefficients entering in Higgs observables at the leading order, using all available Higgs-boson signal strengths from the LHC runs 1 and 2.
Using a prior dependence study of the solutions, we discuss the results within the context of natural-sized Wilson coefficients.
We further study the expected sensitivities to the different Wilson coefficients at various possible future colliders.
Finally, we interpret our results in terms of some minimal composite Higgs models.}

\preprint{\begin{flushright}IFIC/17-48\\ FTUV/18-0305\\FERMILAB-PUB-18-058-T\\ June 2018\end{flushright}}


\begin{document}
\maketitle
\flushbottom


\section{Introduction}
\label{sec:intro}

Within the Standard Model (SM), the Higgs mechanism is essential to understand the mass generation via electroweak symmetry breaking. 
The discovery at the Large Hadron Collider (LHC) of a Higgs-like particle by the ATLAS and CMS experiments~\cite{Aad:2012tfa,Chatrchyan:2012ufa}
could be therefore seen as the experimental confirmation of the last ingredient of the SM. The relevant question in this situation is, however, {\em Is this particle the SM Higgs?}
Indeed, the Higgs is not only a key ingredient of the SM, but also the source of some of the main questions that motivate the belief that there must be new physics beyond the SM.
Such new physics could leave its imprints in deviations of the Higgs couplings with respect to the SM predictions.
The extraordinary performance of the LHC during its eight years of operation has brought us precise measurements of some of these Higgs properties~\cite{Khachatryan:2014kca,Aad:2015mxa,Aad:2015zhl}, 
which can be therefore used to look for this kind of indirect new physics effects.
Also, it is important to note that, with current observables and a precision of at most ten percent in the couplings of the Higgs-like scalar, it is not yet clear if the electroweak symmetry is broken 
by the minimal $SU(2)_L$ doublet of the SM, or if a different mechanism is at work.

The absence of any direct observations of new states 
at the LHC motivates addressing this search of new physics in the Higgs sector in a model-independent way.
The negative findings of direct LHC searches do, in this case, play in our favor, as they suggest
that the scale associated to the new physics is significantly above the electroweak scale. 
This apparent mass gap allows us to take advantage of the techniques of the effective field theories (EFT), to approach the problem in the desired model-independent fashion.
In the EFT formalism, the low-energy effects of high-energy physics are encoded in a set of (non-renormalizable) operators, which are constructed from the field content and the symmetries of the low-energy approximation of the theory. The coefficients of the operators, the Wilson coefficients, depend on the specific UV-completion. In a bottom-up approach, the Wilson coefficients are treated as free parameters, so the EFT describes a large set of different UV-models simultaneously. These techniques became very popular in Higgs physics in the past years \cite{deFlorian:2016spz,Brivio:2017vri}. 

The use of EFTs has several advantages with respect to other model-independent approaches based on phenomenological parameterizations, like, e.g., the so called $\kappa$-framework \cite{LHCHiggsCrossSectionWorkingGroup:2012nn,Heinemeyer:2013tqa}. 
Indeed, the EFT formalism provides a robust theoretical framework where both physical calculations and the connection with UV scenarios can be performed in a well-defined manner. 
On the one hand, EFTs can be connected easily to a given UV-model when the heavy degrees of freedom are integrated out~\cite{delAguila:2000rc,delAguila:2008pw,delAguila:2010mx,Buchalla:2014eca,deBlas:2014mba,Drozd:2015rsp,Shu:2015cxm,delAguila:2016zcb,Henning:2016lyp,Ellis:2016enq,Fuentes-Martin:2016uol,Pich:2016lew,Zhang:2016pja,Dawson:2017vgm,Ellis:2017jns,Criado:2017khh,Rosell:2017kps,Wells:2017vla,deBlas:2017xtg}. On the other hand, the precision of the predictions computed with the effective Lagrangian can be systematically improved by including higher orders of both, the EFT expansion as well as perturbation theory. Moreover, if new states are found in experimental searches, the effective Lagrangian description can be extended to describe these processes too, by including the new particles as part of the low-energy spectrum. (As long as there is, again, a significant mass gap with respect to the next high energy scale). Finally, another asset that comes with the use of EFTs is the power counting information. It gives us an estimated upper limit on the size of the Wilson coefficients. 

Depending on the assumptions of how the Higgs and Goldstone fields are embedded in the low-energy theory, two different EFTs have been developed: the Standard Model EFT (SMEFT) and the Higgs electroweak chiral Lagrangian (\ewXL)\footnote{The original electroweak chiral Lagrangian hypotheses~\cite{Dobado:1989ax,Dobado:1989ue,Dobado:1990zh,Dobado:1990jy,Espriu:1991vm,Herrero:1993nc,Herrero:1994iu} do actually not include a light Higgs in the spectrum. The existence of a physical Higgs, however, can be easily accommodated by extending them adding a Higgs singlet. This is what we will define as our \ewXL\ in sec.~\ref{sec:ewXL}. The same hypotheses have been referred to in the recent literature as the Higgs Effective Field Theory (HEFT)~\cite{Alonso:2015fsp,Brivio:2017vri} or electroweak effective theory~\cite{Pich:2015kwa,Pich:2016lew,Rosell:2017kps}. }.
In the SMEFT, the physical Higgs field and the three Goldstone bosons of electroweak symmetry breaking belong to an $SU(2)_L$ doublet, as part of the low-energy field content. The leading-order Lagrangian of the EFT is the SM one, and the effective expansion is performed in canonical dimensions, i.e.~in inverse powers of the new-physics scale $\Lambda$. In the \ewXL\ the Higgs is treated independently from the Goldstone bosons, reflecting a possible strongly-coupled UV-completion. Because of that, it cannot be expanded in canonical dimensions and chiral dimensions have to be used instead \cite{Buchalla:2013eza}. The leading-order Lagrangian is more general than in the SM, focussing on new physics in the Higgs sector and testing the SM-Higgs hypothesis at the same time. One of the main points of this article
is to update our current knowledge of the \ewXL, using the latest Higgs-boson results from the LHC run 2, and also to extrapolate these results to what we could learn at future Higgs factories.

Both the SMEFT and the \ewXL~have been used in fits to experimental data \cite{Han:2004az,delAguila:2011zs,Ciuchini:2013pca,deBlas:2013qqa,deBlas:2013gla,Pomarol:2013zra,deBlas:2014ula,Falkowski:2014tna,Corbett:2015ksa,Buckley:2015nca,deBlas:2015aea,Falkowski:2015jaa,Berthier:2015gja,Buchalla:2015qju,Englert:2015hrx,Falkowski:2015krw,Butter:2016cvz,Brivio:2016fzo,deBlas:2016ojx,deBlas:2016nqo,Falkowski:2017pss,Englert:2017aqb,deBlas:2017wmn,Jana:2017hqg} to search for indirect signals of new physics. 
Apart from the experimental information, it is sometimes also useful to take theory considerations in such analyses into account. 
In particular, we are interested here in testing the \ewXL~hypotheses within the regime that is expected by the power-counting rules of the chiral Lagrangian, or, conversely, to clarify to what extent current data is sensitive to ``natural''-sized EFT contributions. 
This can be easily done within the framework of Bayesian statistics, where such theoretical information can feed into priors in the process of parameter estimation \cite{Wesolowski:2015fqa}.
In Bayesian inference (see for example \cite{D'Agostini:2003qr} for a review), probabilities are interpreted as degree-of-belief and always depend on some background information. The central formula of Bayesian inference can then be obtained from the Bayes Theorem, 
\begin{align}
  \begin{aligned}
    \label{eq:Bayesian}
    \prob{\text{hypothesis}}{\text{data},I} &= \frac{\prob{\text{data}}{\text{hypothesis},I} \cdot \prob{\text{hypothesis}}{I}}{\prob{\text{data}}{I}}\\
     &\Leftrightarrow\\
     posterior &\sim  likelihood \times prior,
  \end{aligned}
\end{align}
where we denote the probability of an event $X$, given the background information $I$ as $\prob{X}{I}$. 
In this formalism, the relevant information to address the questions \emph{``Do we see indirect signs of new physics in current Higgs data?''} or, conversely, 
 \emph{``How large can new physics be, while still being consistent with Higgs data?''}, 
 is contained in the posterior distributions for the Wilson coefficients.
 These directly feed from the information of the likelihood, where we included the most recent experimental data.
Bayesian methods are now widely-used in Higgs fits~\cite{Fichet:2012sn,Dumont:2013wma,deBlas:2014ula,Bergstrom:2014vla,Buchalla:2015qju,deBlas:2017wmn}, and it is also the framework that we will use in our EFT studies, so we can discuss the consistency of the experimental results with the EFT considerations.

The paper is organized as follows.
In section~\ref{sec:ewXL} we review the basics of the electroweak chiral Lagrangian, and present the actual parameterization we will be testing in our fits. 
These are performed using the {\tt HEPfit} package \cite{hepfit,hepfitsite}, in which we implement the relevant signal strengths computed with the \ewXL. 
We discuss {\tt HEPfit}, the settings of the fits, and the experimental data set included in our analyses in section~\ref{sec:Methodology}. Section~\ref{sec:fits_pheno} contains the main phenomenological results of this article. We discuss the constraining power of current data and the impact of different priors in testing the \ewXL~power counting. We present the final result of our fit, describing the uncertainties and correlations of the \ewXL~parameters. 
These results are extended in section~\ref{sec:projections}, with the study of the projected uncertainties of the Wilson coefficients at future colliders. Both current and future results are then related in section~\ref{sec:CHM} to the $SO(5)/SO(4)$ minimal composite Higgs model. We conclude in section~\ref{sec:conclusions}. Supplementary information is presented in three appendices. In appendix~\ref{app:ThExpr} we list the relevant formulas for the calculation of the Higgs signal strengths. We also compare the results of the \ewXL~fit with those obtained within the phenomenological approach provided by the $\kappa$-formalism~\cite{LHCHiggsCrossSectionWorkingGroup:2012nn,Heinemeyer:2013tqa} in appendix~\ref{app:kappa}. Finally, we collect some of the input used in the fits presented in Section~\ref{sec:projections} in appendix~\ref{app:futcol_data}.


\section{The Higgs electroweak chiral Lagrangian}
\label{sec:ewXL}

In this paper we use the bottom-up EFT that was derived in \cite{Buchalla:2015wfa}, to describe potentially large deviations from the SM in Higgs observables. It is based on the electroweak chiral Lagrangian \cite{Dobado:1989ax,Dobado:1989ue,Dobado:1990zh,Dobado:1990jy,Espriu:1991vm,Herrero:1993nc,Herrero:1994iu,Feruglio:1992wf,Bagger:1993zf,Koulovassilopoulos:1993pw,Burgess:1999ha,Wang:2006im,Grinstein:2007iv,Azatov:2012bz,Alonso:2012px,Buchalla:2012qq,Buchalla:2013rka,Buchalla:2013eza}.
Such large deviations in the Higgs sector are motivated in many scenarios of physics beyond the SM, like for example composite Higgs models \cite{Agashe:2004rs,Contino:2006qr,Contino:2010rs,Carena:2014ria}. 
As any bottom-up EFT, its Lagrangian is completely defined via the particle content and the symmetries of the low-energy theory, while the effective expansion is determined by power counting rules. The explicit assumptions that go into the construction of the specific EFT we consider are described in what follows:
\begin{itemize}
\item {\bf Particles:} We assume the SM particle content, but no relation between the Higgs scalar $h$ and the three Goldstone bosons $\varphi_{i}$ of electroweak symmetry breaking.
\item {\bf Symmetries:} We assume the SM gauge symmetry, $$SU(3)_{C}\times SU(2)_{L}\times U(1)_{Y}\rightarrow SU(3)_{C}\times U(1)_{Q},$$ and that the new physics conserves custodial symmetry. Therefore, the global symmetry breaking pattern in the scalar sector is $$SU(2)_{L}\times SU(2)_{R} \rightarrow SU(2)_{L+R}.$$ We further assume conservation of baryon and lepton number, a SM-like flavour structure in the Yukawa interactions of the Higgs, as well as $\mathcal{CP}$-symmetry in the Higgs sector. The latter is also motivated by current experimental constraints~\cite{Dekens:2013zca,Cirigliano:2016njn,Cirigliano:2016nyn,Aaboud:2018xdt}. 
\item The {\bf power counting} of the electroweak chiral Lagrangian is given by a loop expansion, which equivalently can be expressed in terms of chiral dimensions \cite{Buchalla:2013eza}. With the assignments
$$[\text{bosons}]_{\chi} = 0$$ 
and $$[\text{fermion bilinears}]_{\chi} = [\text{derivatives}]_{\chi} =[\text{weak couplings}]_{\chi}\linebreak=1,$$ the total chiral dimension of a term in the Lagrangian equals $2L+2$, with $L$ being the loop order and therefore the order of the EFT expansion.

If the new physics is decoupled from the SM to some degree, it is useful to parametrize the deviations from the SM by the parameter $\xi = v^{2}/f^{2}$, where $v\approx 246$ GeV is the electroweak vacuum expectation value, and $f$ is the scale of new physics. The latter could correspond, for example, to the scale of global symmetry breaking in composite Higgs models. If $\xi\ll 1$, we can perform an expansion in $\xi$ (and therefore in canonical dimensions) on top of the loop expansion. This yields a double expansion in $\xi$ and $1/16\pi^{2}$ \cite{Buchalla:2014eca}. 
\end{itemize}

The leading-order chiral Lagrangian, not expanded in $\xi$ (i.e.~for $\xi = \mathcal{O}(1)$), is then \cite{Buchalla:2013rka}
\begin{align}
  \begin{aligned}
    \label{eq:1}
    \mathcal{L}_{\text{LO}} &= -\frac{1}{2} \langle G_{\mu\nu}G^{\mu\nu}\rangle -\frac{1}{2}\langle W_{\mu\nu}W^{\mu\nu}\rangle -\frac{1}{4} B_{\mu\nu}B^{\mu\nu} \\
  &+i\bar{q}_{L}\slashed{D}q_{L} +i\bar{\ell}_{L}\slashed{D}\ell_{L} +i\bar{u}_{R}\slashed{D}u_{R} +i\bar{d}_{R}\slashed{D}d_{R} +i\bar{e}_{R}\slashed{D}e_{R} \\
  &+\frac{v^2}{4}\ \langle D_\mu U^\dagger D^\mu U\rangle \left( 1+F_U(h)\right) +\frac{1}{2} \partial_\mu h \partial^\mu h - V(h)\\
  &- \frac{v}{\sqrt{2}} \left[ \bar{q}_{L}  Y_u(h)  U P_{+}q_{R} + \bar{q}_{L}  Y_d(h) U P_{-}q_{R}  + \bar{\ell}_{L}  Y_e(h) U P_{-}\ell_{R} + \text{ h.c.}\right] , 
  \end{aligned}
\end{align}
where $U=\exp{(2i\varphi_{a}T^{a}/v)} $ collects the Goldstone bosons, $T^{a}$ are the generators of $SU(2)$, $P_{\pm}=1/2 \pm T^{3}$, and $\langle \cdot \rangle$ denotes the trace.  
As already said, we do not assume a relation between $h$ and the Goldstone bosons in $U$. This yields free coefficients for all Higgs couplings in $V(h), F_{U}(h)$, and $Y_{\psi}(h)$ for any fermion $\psi$. To allow for a possible strongly-coupled origin of $h$, we do not truncate the polynomials at any order in $h$. 

All terms in $\mathcal{L}_{\text{LO}}$ have chiral dimension two. The list of NLO operators (i.e.~with chiral dimension four) is lengthy \cite{Buchalla:2013rka} and we will not list all the operators here, as only a few operators are important for our analysis.
In particular, we will focus on single-Higgs production processes. (We briefly comment on double-Higgs production at the end of this section.)
Working at the leading order in each process we can therefore focus on operators with a single Higgs field. At tree level, this includes couplings of $h$ to $W^{+}W^{-}$, $ZZ$, $\bar{t}t$, $\bar{b}b$, $\bar{c}c$, $\tau^{+}\tau^{-}$, and $\mu^{+}\mu^{-}$. Normalized to their SM values, we expect the couplings to be $1\pm \mathcal{O}(\xi)$ by the power counting arguments from above. Couplings to lighter fermions have not been observed so far, and therefore we do not include them in our fit. Nevertheless, to illustrate what effect those couplings could have in the fit, we still include the effective charm coupling. 

\begin{figure}[t]
\minipage{0.32\textwidth}
\begin{overpic}[width=\linewidth]{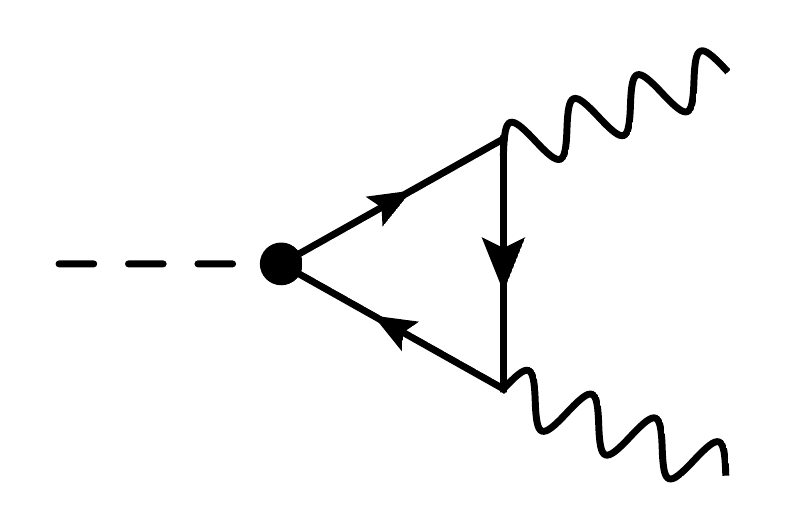}
\put(0,32){$h$}
\put(95,55){$\gamma$}
\put(95,10){$\gamma$}
\end{overpic}
\endminipage\hfill
\minipage{0.32\textwidth}
\begin{overpic}[width=\linewidth]{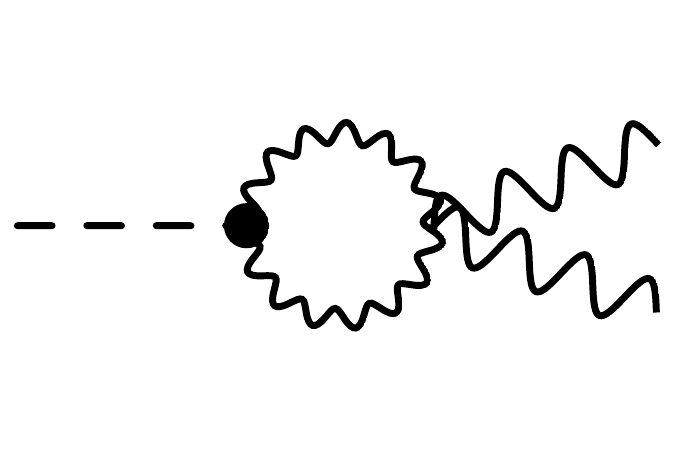}
\put(-5,32){$h$}
\put(100,50){$\gamma$}
\put(100,15){$\gamma$}
\end{overpic}
\endminipage\hfill
\minipage{0.32\textwidth}%
  \begin{overpic}[width=\linewidth]{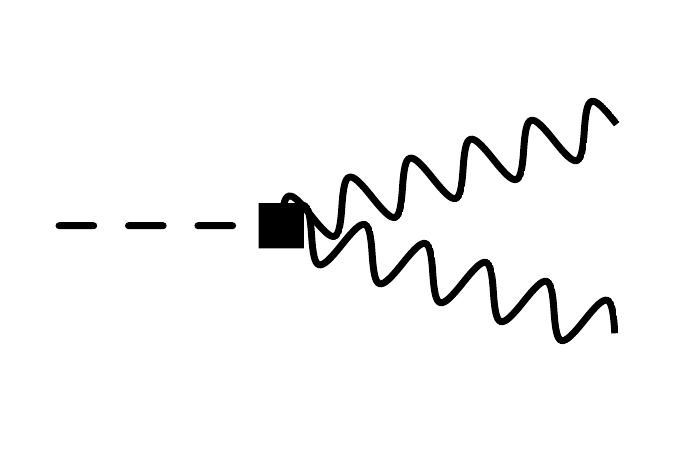}
\put(0,32){$h$}
\put(95,50){$\gamma$}
\put(95,15){$\gamma$}
\end{overpic}
  \endminipage
  \caption{Schematic contributions to the $h\rightarrow\gamma\gamma$ amplitude. Black dots denote vertices from $\mathcal{L}_{\text{LO}}$, black squares from $\mathcal{L}_{\text{NLO}}$. All of these contributions introduce deviations from the SM at $\mathcal{O}(\xi/16\pi^{2})$.}\label{fig:1loop}
\end{figure}
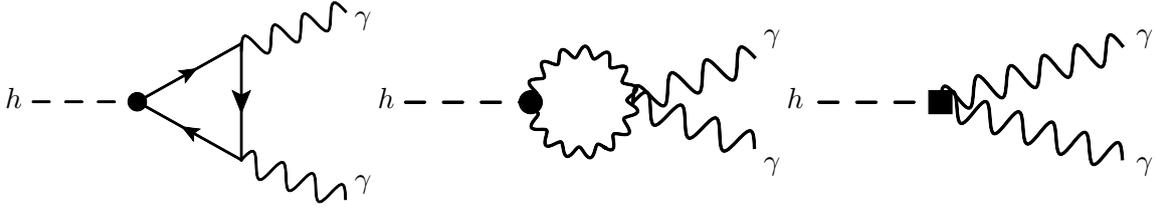

There is also experimental information for loop-induced processes, involving Higgs couplings to $gg, \gamma\gamma, $ and $Z\gamma$. The amplitudes of these processes receive contributions of $\mathcal{O}((1+\xi)/16\pi^{2})$ coming from the modified leading-order couplings that enter in the loop. In addition, there are operators in $\mathcal{L}_{\text{NLO}}$ that, when included at tree level, contribute at $\mathcal{O}(\xi/16\pi^{2})$ in these amplitudes. We therefore include these operators as well. Figure~\ref{fig:1loop} shows the contributions of $\mathcal{L}_{\text{LO}}$ and $\mathcal{L}_{\text{NLO}}$ to the example process $h\rightarrow \gamma\gamma$ schematically.

The resulting Lagrangian, which we use for our fit, is then \cite{Buchalla:2015qju,Buchalla:2015wfa}\footnote{Similar parametrizations have been also discussed, using phenomenological motivations, in Refs.~\cite{Carmi:2012yp,Espinosa:2012im,Giardino:2013bma,Ellis:2013lra,Einhorn:2013tja,Bernon:2014vta,Flament:2015wra}.}
\begin{align}
  \begin{aligned}
    \label{eq:2}
    \mathcal{L}_{\text{fit}} &= 2 c_{V} \left(m_{W}^{2}W_{\mu}^{+}W^{-\mu} +\tfrac{1}{2} m^2_Z Z_{\mu}Z^{\mu}\right) \dfrac{h}{v} - \sum_{\psi}c_{\psi} m_{\psi} \bar{\psi} \psi \dfrac{h}{v} \\
 &+ \dfrac{e^{2}}{16\pi^{2}} c_{\gamma} F_{\mu\nu}F^{\mu\nu} \dfrac{h}{v}+ \dfrac{e^{2}}{16\pi^{2}} c_{Z\gamma} Z_{\mu\nu}F^{\mu\nu} \dfrac{h}{v}+\dfrac{g_{s}^{2}}{16\pi^{2}} c_{g}\langle G_{\mu\nu}G^{\mu\nu}\rangle\dfrac{h}{v},
  \end{aligned}
\end{align}
where the Wilson coefficients are\footnote{While the assumptions that lead to these generic power counting estimates hold for many models of new physics, there are also exceptions: if there are, for example, different sources of electroweak symmetry breaking for different generations of fermions, it could be that the Higgs couplings to the light fermions are enhanced by larger factors, see \cite{Altmannshofer:2015esa}.}
\begin{equation}
  \label{eq:5}
    c_{i} =  c_{i}^\SM + \mathcal{O}(\xi),
\end{equation}
\begin{equation}
  \label{eq:cSM}
    c_{i}^\SM = \begin{cases} 1 & \text{for } i = V, t, b, c, \tau, \mu\\ 0 & \text{for } i = g, \gamma, Z\gamma. \end{cases}
\end{equation}
Note that the coefficients $c_{\gamma}$ and $c_{Z\gamma}$ are independent at the considered order. These can be induced by the following three operators,
\begin{align}
  \begin{aligned}
    \label{eq:3}
    \mathcal{O}_{Xh1} &=g^{\prime 2} B_{\mu\nu} B^{\mu\nu}\, F_{Xh1}(h), \\
    \mathcal{O}_{Xh2} &=g^{2} \langle W_{\mu\nu} W^{\mu\nu}\rangle\, F_{Xh2}(h), \\
    \mathcal{O}_{XU1} &=g' g B_{\mu\nu}\langle W^{\mu\nu} U T_3 U^\dagger\rangle\, (1+F_{XU1}(h)).
  \end{aligned}
\end{align}
These operators contribute to the following four interactions
\begin{align}
  \begin{aligned}
    \label{eq:4}
    hF_{\mu\nu}F^{\mu\nu}, \; hZ_{\mu\nu}F^{\mu\nu}, \;hZ_{\mu\nu}Z^{\mu\nu}, \;hW_{\mu\nu}W^{\mu\nu},
  \end{aligned}
\end{align}
yielding one linear dependent operator in~\eqref{eq:4}. However, corrections induced by the two last operators are subleading (at $\mathcal{O}(\xi/16\pi^{2})$) compared to the leading-order contributions parametrized by $c_{V}$ (at $\mathcal{O}(\xi)$) and are therefore neglected. 

As indicated above, we focus our study to single-Higgs processes. 
To describe double-Higgs production consistently within the electroweak chiral Lagrangian, we would need to include several more parameters in the fit (at least three more to describe gluon fusion, corresponding to the interactions $h^{3},\bar{t}th^{2},ggh^{2}$~\cite{Grober:2015cwa,deFlorian:2016spz,Kim:2018uty}). Given the low current sensitivity of the ATLAS and CMS experiments to double-Higgs production (the best upper limits are of the order of 20-30 times the SM~\cite{ATLAS:2016ixk,CMS:2017ihs}) and that these parameters cannot be constrained by the other available measurements, we decided to not include double-Higgs production in our analysis. 

Finally, let us mention that the analysis with the leading-order electroweak chiral Lagrangian is closely related \cite{Buchalla:2015wfa}, but not identical to the $\kappa$-framework \cite{LHCHiggsCrossSectionWorkingGroup:2012nn,Heinemeyer:2013tqa}, which was introduced by the LHC Higgs cross section working group. We discuss the relation and the differences in appendix~\ref{app:kappa}.


\section{Methodology}
\label{sec:Methodology}
In this section we review the details of our phenomenological analysis.  
The fits are performed using the {\tt HEPfit} package \cite{hepfit,hepfitsite}, a general tool designed to easily combine the information from direct and indirect searches and test the SM and extensions including new physics effects. The code is available under the GNU General Public License. The current developers' version can be downloaded at \cite{hepfitGit}. The flexibility of the {\tt HEPfit} framework allows to easily introduce new physics models and observables as external modules to the main core of the code, which we have used to implement the Higgs electroweak chiral Lagrangian (see section~\ref{sec:ewXL}) and its modifications on the Higgs sector. {\tt HEPfit} includes a Markov-Chain Monte-Carlo implementation provided by the Bayesian Analysis Toolkit \cite{Caldwell:2008fw}, which we use to perform a Bayesian statistical analysis of the model. More details on the code can be found in \cite{hepfitsite}, and related phenomenological studies using it are presented in \cite{deBlas:2016ojx,Chowdhury:2017aav}. As an illustration of the performance of {\tt HEPfit}, the global fit presented later in this paper ($9$ parameters ---using both Gaussian and flat priors--- and a total of $126$ observables in the likelihood) was performed using a total of $24$ Markov chains with $10^{7}$ iterations each. This can be done within a total of 1200 CPU hours, or roughly two days if parallelized over 24 cores.

On the experimental side, we include in the likelihood all available Higgs boson signal strengths measured by ATLAS and CMS, both at the LHC runs 1 and 2, as well as the experimental results obtained by the CDF and D$\cancel{0}$ collaborations at the Tevatron. The actual experimental results included in the fits are summarized in table~\ref{tab:signal_strengths}, which contains the references to the experimental measurements used for each Higgs decay channel. Whenever available, we include in the likelihood directly the signal strengths measured per category in each experimental analysis. The signal strengths are defined by
\begin{equation}
\mu=\sum_i w_i r_i,~\mbox{ with }~r_i=\frac{(\sigma\times \text{Br})_i}{(\sigma_{\rm{SM}}\times \text{Br}_{\rm{SM}})_i}~\mbox{ and }~w_i=\frac{\varepsilon_i(\sigma_{\rm{SM}}\times \text{Br}_{\rm{SM}})_i}{\sum_j\varepsilon_j(\sigma_{\rm{SM}}\times \text{Br}_{\rm{SM}})_j},
\end{equation}
with the sums running over all the different production mechanisms that contribute to the study of each final state. The SM predictions for the different production cross sections and branching ratios are taken from \cite{Heinemeyer:2013tqa}. 
The experimental efficiencies, $\varepsilon_i$, are assumed to be SM-like:  $\varepsilon_i\approx \varepsilon_i^{\rm SM}$. This is a good approximation in presence of small new physics effects, or if these do not modify significantly the kinematical distributions of the final states. The validity of this approximation must nevertheless be checked a posteriori, in light of the results of the fit pertaining the new physics effects. In any case, none of the interactions considered here introduce vertices with tensor structures different than the SM ones and based on the natural size expected for the EFT coefficients 
one expects $\varepsilon_i\approx \varepsilon_i^{\rm SM}$ to be a good approximation.  
Correlations between the observed signal strengths for the different categories are usually not provided by the experimental groups and are ignored here.\footnote{This also includes the theory correlations between the observed rates. See~\cite{Fichet:2015xla,Arbey:2016kqi} for a detailed treatment of such theory uncertainties and correlations.} In some cases, the references for the experimental analyses do not provide all the information needed to reconstruct the signal strengths per categories. In such cases we use as observables the fit values of the signal strength per production mechanism (e.g.~$\mu_{ggF,~\!VBF,~\! Vh,~\! tth}$) or decay channel. 
Finally, for the $h\to Z \gamma$ channel, only limits on the signal strengths are available. In that case the limit is transformed into a Gaussian contribution to the likelihood. (In any case, the constraining power of the $h\to Z \gamma$ channel is very small and, within the model-independent hypotheses we are testing, has almost no impact on the fits.)

\begin{table}[t!]
{\small
\begin{center}
\begin{tabular}{ c| c c c c| c c c c }
\toprule
Channel&\multicolumn{4}{c|}{ATLAS}&\multicolumn{4}{c}{CMS}\\
                                    &    7/8 TeV     & $L$[fb$^{-1}$]   &    13 TeV   & $L$[fb$^{-1}$]  &     7/8 TeV   & $L$[fb$^{-1}$]  &    13 TeV  & $L$[fb$^{-1}$]  \\         
\midrule
$h \to \gamma\gamma$&\cite{Aad:2014eha}&$24.8$&\cite{Aaboud:2018xdt}&$36.1$&\cite{Khachatryan:2014ira}&$24.8$&\cite{CMS:2017rli}&$35.9$ \\
\midrule
\multirow{2}{*}{$h \to Z Z$}&\cite{Aad:2014eva}&$24.8$ & \cite{Aaboud:2017vzb}&$36.1$ &\cite{Khachatryan:2014jba}&$24.8$ &\cite{Sirunyan:2017exp}&$35.9$ \\
           && &\cite{Aaboud:2017jvq}&$36.1$&& &\cite{Sirunyan:2018shy}&35.9  \\ 
\midrule
\multirow{3}{*}{$h \to W W$}&\cite{ATLAS:2014aga,Aad:2015ona}&$24.8$ &\cite{ATLAS:2016gld}&$5.8$ &\cite{Chatrchyan:2013iaa}&$24.3$ &\cite{Sirunyan:2018egh}&$35.9$  \\ 
           && &\cite{Aaboud:2017jvq}&$36.1$&& &\cite{Sirunyan:2018shy}&35.9  \\ 
           && &\cite{ATLAS-CONF-2018-004}&$36.1$&& && \\ 
\midrule
\multirow{3}{*}{$h \to \bar{b}b$}&\cite{Aad:2014xzb}&$25$ &\cite{Aaboud:2017rss} &$36.1$ &\cite{Chatrchyan:2013zna}&$24$ &\cite{CMS:2016mmc}&$2.3$ \\
&\cite{Aad:2015gra}&$20.3$ & \cite{Aaboud:2017xsd}&$36.1$ &\cite{Khachatryan:2014qaa}&$24.8$ &\cite{Sirunyan:2018mvw,Sirunyan:2018ygk}&$35.9$ \\
&& & & && &\cite{Sirunyan:2017elk}&$35.9$ \\
\midrule
$h \to \tau^+\tau^-$&\cite{Aad:2015vsa}&$24.8$ &\cite{Aaboud:2017jvq}&$36.1$&\cite{Chatrchyan:2014nva}&$24.6$&\cite{Sirunyan:2017khh,Sirunyan:2018shy}&$35.9$\\
\midrule
$h \to \mu^+ \mu^-$&\cite{Khachatryan:2016vau}&$25$ &\cite{Aaboud:2017ojs}&$36.1$ &\cite{Khachatryan:2016vau}&$25$&\cite{CMS-PAS-HIG-17-019}&$35.9$ \\
\midrule
$h \to Z\gamma$&\cite{Aad:2015gba}&$25$ &\cite{Aaboud:2017uhw}&$36.1$ &\cite{Chatrchyan:2013vaa}&$24.6$ &\cite{CMS-PAS-HIG-17-007}& 35.9  \\
\bottomrule
\end{tabular}
\caption{
Higgs boson signal strengths included in our fits to the electroweak chiral Lagrangian, classified according to the final states and indicating the integrated luminosity, $L$, of the corresponding data sets. Multiple entries in a single cell refer to different production modes. We summarize 7 and 8 TeV data for brevity. We also include CDF~\cite{Aaltonen:2013ipa} and D\cancel{0}~\cite{Abazov:2013gmz} data sets in our fit.
\label{tab:signal_strengths}}
\end{center}
}
\end{table}

On the theory side we consider the leading order (LO) corrections from eq.~\eqref{eq:2} to the SM Higgs boson production cross sections and branching ratios. 
The explicit expressions for the different observables are provided in appendix~\ref{app:ThExpr}.

In our analysis we work within a Bayesian statistical framework, where the information we know about the model parameters before the analysis can be encoded into a prior distribution. 
In this regard, all the SM parameters are taken as fixed parameters.
(In the expressions for the new physics corrections in appendix~\ref{app:ThExpr}, the SM inputs have been fixed to the central values from the fit in \cite{deBlas:2016ojx}.) 
For the coefficients of the EFT, this a priori information comes from the EFT power counting: $c_{i}= c_{i}^{\text{SM}}+\mathcal{O}(\xi)$, see eq.~\eqref{eq:5}. To decide how to implement this information into a prior we follow the \emph{principle of maximum entropy}~\cite{Jaynes:1957zza}, which selects the prior that reflects the current state of knowledge best \cite{Schindler:2008fh}. The use of a flat prior only contains information in the boundaries of the region where it is non-vanishing but, apart from that, it does not reflect a preference for a particular size of the model parameters. This is convenient to show (exclusively) all the information from the actual data included in the likelihood in the desired region. 
A Gaussian prior, on the other hand, seems to be more convenient to implement the information in eq.~\eqref{eq:5}, by choosing the SM expectation of the coefficients as a mean value and adjusting the standard deviation, $\sigma$, of the distribution to favour (penalize) effects within (beyond) the expected size of $\mathcal{O}(\xi)$.
To a certain extent, the result of the fit is prior-dependent. As we will see in sec.~\ref{sec:fits_flat}, the result of the fit changes when we restrict the Wilson coefficients $c_{i}$ to be around the SM solution only. However, the fit results should \emph{not} depend on the form we put the ``condition of natural-sized coefficients'' into equations. This is why we study the prior dependence and its impact on the quality/consistency of our results using both, flat and Gaussian distributions. 

The information in the prior can also help to prevent from \emph{overfitting} when fitting a theoretical model to experimental data~\cite{Wesolowski:2015fqa}. 
The term \emph{overfitting} refers to cases in which the parameter values are very fine-tuned to data. As we will explain, in our EFT, this would for example correspond to an unnaturally large value of the Higgs-charm coupling. A reasonable choice of priors, ensuring the condition of ``natural-sized Wilson coefficients in the EFT'', reduces the risk of \emph{overfitting} substantially~\cite{Wesolowski:2015fqa}.


\section{Phenomenological results}
\label{sec:fits_pheno}

As mentioned in the previous section, the use of flat priors is convenient to establish the constraining power of the experimental data contained in the likelihood in absence of extra information about the model. We therefore start by presenting our results using flat priors for all EFT coefficients, and then move to check the stability of such results under the hypothesis of natural-sized Wilson coefficients. The main results presented here are obtained with the full data from Tevatron and LHC runs 1 and 2 (see table~\ref{tab:signal_strengths}). We will also comment in section~\ref{sec:SMsolution} on the comparison of the constraining power of run 1 and run 2 data separately.


\subsection{Flat priors}
\label{sec:fits_flat}

From the equations of the production cross sections and decay widths in appendix~\ref{app:ThExpr} the existence of several approximate symmetries in the EFT parameter space is apparent. This will yield a certain degree of degeneracy in the posterior distributions. Indeed, all observables are unchanged under a simultaneous change of sign 
\begin{equation}
c_i\rightarrow -c_i,~~~\forall i.
\label{eq:cisign}
\end{equation}
This follows from a more general reparameterization invariance of the electroweak chiral Lagrangian where each Higgs couplings is modified as
\begin{equation}
c_i\rightarrow (-1)^{n_h} c_i,~~~\forall i,
\end{equation}
with $n_h$ the number of Higgs fields in each Lagrangian term. This change of sign can then be absorbed in the Higgs field, $h\to -h$, leaving the action invariant.

The invariance \eqref{eq:cisign} is enhanced at the tree-level, since observables remain unchanged under $c_i\rightarrow -c_i$ for each coefficient independently. 
Only when we include loop-generated observables, e.g.~$gg\to h$ or $h\to \gamma\gamma$, the interplay between the different diagrams contributing to the effective $hgg$, 
$h\gamma\gamma$ and $hZ\gamma$ vertices makes the fit sensitive to the relative sign of the coefficients. The description of loop-generated observables, however,
also introduces three extra parameters that do not enter in the tree-level decay widths, namely $c_{g}$, $c_{\gamma}$ and $c_{Z\gamma}$, which each enter only in one effective vertex.
In general, for a given point in the $c_i$ parameter space, one can therefore flip the signs of $c_\psi$ and $c_V$ independently, and adjust the previous local parameters to obtain exactly the same prediction as the original point, and therefore the same likelihood. 

Note now that there is basically no direct sensitivity in current data to the charm coupling (e.g.~via $h\to \bar{c}c$) or any of the other light quarks. (The most stringent bound limits the $h\to \bar{c}c$ signal strength to be smaller than $110$ \cite{Aaboud:2018fhh}.)
From the point of view of the fit, the corresponding Wilson coefficients can therefore be used as compensating parameters to balance out the effects of deviations of the other $c_i$ away from the SM. Firstly, the absence of a significant handle to $\mathrm{Br}(h\to \bar{q}q)$ would allow to use $c_q\gg1$ to effectively compensate a global enhancement of the decay widths for all the observed channels, leaving the corresponding branching ratios intact. 
Secondly, $c_q$ could play a role similar to $c_{g}$, $c_{\gamma}$ and $c_{Z\gamma}$, and be adjusted to cancel some of the contributions induced by the other couplings in all the one-loop effective vertices. (The couplings to light leptons, on the other hand, could only play a similar role in electroweak loops.) 
Note that these partial cancellations in the effective loop vertices are possible for several different patterns of deviations of the EFT couplings. However, since the SM branching ratios into light quarks are very small, the first effect (compensating enhancements in the observed decay widths) is only possible for couplings that are larger, in magnitude, than in the SM. 
Both effects also require very large values of the corresponding $c_q$. For the case of the charm, being the heaviest quark after the bottom, $c_c$ can still have a visible effect for ${\cal O}(1-10)$ values.~\footnote{While weak, the experimental limits in the $h\to \mu\mu$ channel are still restrictive enough to prevent any compensating effect of the $c_\mu$ in electroweak loops.}
Let us illustrate this with a simple example. If we switch off the one-loop parameters ($c_g=c_\gamma=c_{Z\gamma}=0$), enhance the $c_{c}$ coupling by a factor of $5$ and set the other tree-level couplings to $c_V=c_t=c_b=c_\tau=c_\mu=1.25$, we will increase the total decay width of the Higgs by almost a factor of $2.3$.~\footnote{
In this regard, let us comment that the existing bounds on the Higgs width, e.g.~\cite{Khachatryan:2016ctc}, do not apply in a straightforward manner to our analysis. Indeed, current experimental limits on $\Gamma_h$ depend on certain theory assumptions, like that gluon fusion production is dominated by the effects of the top loops, while the \ewXL~hypotheses also allow extra contributions coming from $c_g$.
It is because of this that we ignore such bounds on $\Gamma_h$ in our fits. 
In any case, while the strongest experimental bounds ---$\Gamma_h<13$ MeV at $95\%$ C.L.~\cite{Khachatryan:2016ctc}--- could alleviate the overfitting issue by preventing excessively large values of $c_c$, it does not completely avoid it. The example discussed is, in fact, not only consistent with the above mentioned limit on $h\to \bar{c}c$, but also with $\Gamma_h<13$ MeV. Disregarding the width measurements completely, much larger values of $c_{c}$ are possible, with the other $c_{i}$ adjusted accordingly.}
With this knowledge, we can estimate the signal strengths for tree-level processes. For instance, the $h\to VV$ signal strengths in the VBF or Vh production channels roughly scale with $c_V^4/2.3\approx 1.06$ now. But also signals strengths involving loop processes like gluon fusion or the diphoton decay do not feature strong deviations from the SM values. The charm contribution to the loop functions is about 1\% of the top terms and its relative boost by a factor of $5/1.25=4$ is not sufficient to contribute in a significant way to the total amplitude. Even if the experimental resolution was precise enough to measure these effects in the loop observables, we would always have the freedom to compensate the charm loop by the local terms. With a little bit of tuning, it is easy to bring all signal strengths to SM values within the desired accuracy. 

The fact that $c_c$ is a priori experimentally unconstrained with current data, together with the above-mentioned possibility of being used to partially cancel the effects of other couplings, introduces a problem in the global fits if we treat it as a floating parameter. 
Assuming unbounded flat priors for all $c_i$ parameters, such a fit would suffer from a clear case of overfitting of the EFT hypothesis to data, 
in the sense that the EFT parameters are ``pulled'' towards unnatural values while preserving the quality of the fit to data. 
As illustrated in the example above, large absolute values of $c_c$ offer more room for possible cancellations between this and the other parameters. 
This opens the regions of the parameter space where large values of $c_c$ and tuned combinations of the other parameters leave the production cross sections times branching ratios approximately unchanged with respect to the SM expectation, and are therefore consistent with the experimental data. 
Since the likelihood along these regions is approximately the same as the SM one, we would expect some degeneracy in the posterior for the corresponding parameters, in correlation with $c_c$. 

Instead, what we observe is a preference for scenarios with $|c_c| \gg 1$ and absolute values for the other $c_i$ above the SM expectation. 
This can be understood as follows: around the SM limit, all $c_i$ must have very specific values in order to obtain SM-like signal strengths. If, on the other hand, $|c_c|$ is large, the other parameters have more freedom to compensate this contribution with various different correlations. In other words, if we do not know anything about $c_c$, the larger the $c_c$ is the more fine-tuned the SM point looks like, and it is therefore less likely to be scanned by the MCMC, i.e.~the SM neighborhood seems less probable. 
Once again, let us illustrate this with another example: the most precise measurement of a Higgs decaying to a $\tau$ pair leaves an uncertainty of around 35\%. Leaving all other observables like in the SM, we can vary $c_\tau$ only between 0.81 and 1.16. But if the total width is scaled by 2.3 and the production enhanced by $1.25^2$, the possible range for $c_\tau$ is between 0.98 and 1.41 and thus by a factor $\approx 1.25$ larger.
In general, if we allow all parameters $c_i$ to vary independently, large values of $c_c$ induce a global suppression in all signal strengths, via an enhancement of the total decay width with respect to the SM. This allows a wider range of different parameters to be consistent with data 
than in the case of a SM-sized decay width.
The larger size of the allowed regions together with the multiplicity of the solutions thus leads to a preference for the large $c_c$ values.

\begin{figure}[t]
   \begin{picture}(400,150)(0,0)
      \put(0,0){\includegraphics[width=400pt]{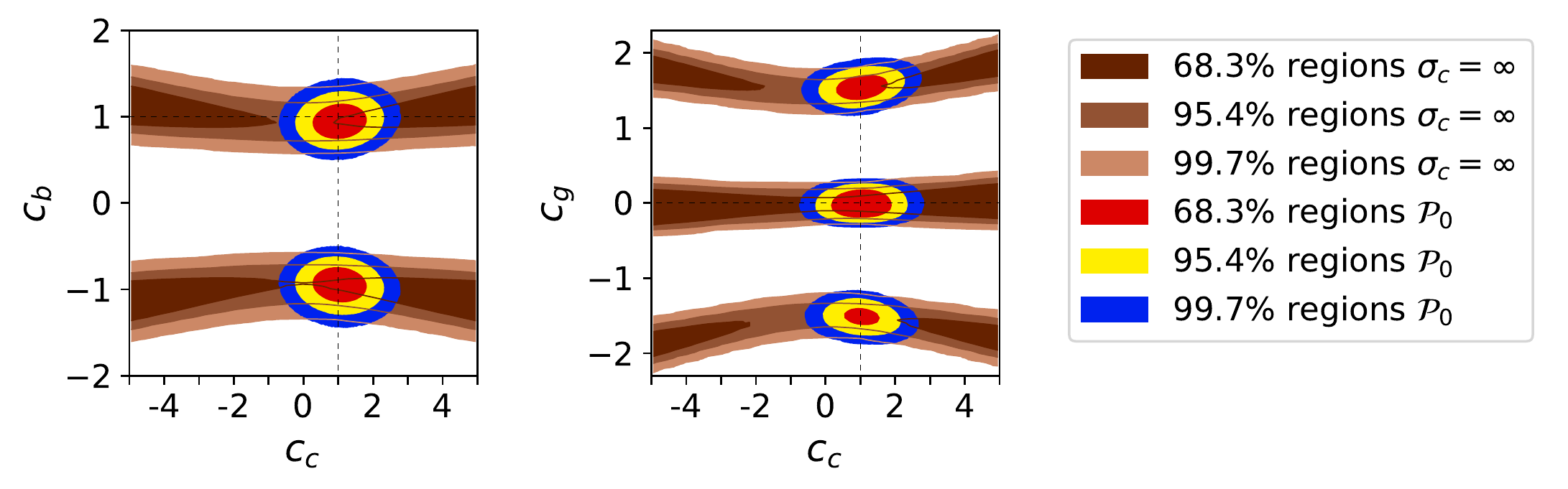}}
      \put(270,13){\includegraphics[width=50pt]{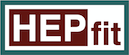}}
   \end{picture}
   \caption{Constraints on the $c_b$ vs.~$c_c$ (left) and $c_g$ vs.~$c_c$ (right) planes obtained from a fit using a flat prior ($\sigma_c=\infty$) within $[-5,5]$ for $c_c$ (brown regions) and the same assuming a Gaussian prior for $c_c$ with $\sigma_c=0.5$. Both fits use the same experimental data.
While the fit with a Gaussian prior for $c_c$ is concentrated around the SM limit, denoted by the black dashed lines, in the flat case the posterior for all parameters is significantly broadened  and also shows a tendency towards larger values for all parameters $c_i$. }
   \label{fig:Overfitting}
\end{figure}

This overfitting issue for $c_c$ is illustrated in fig.~\ref{fig:Overfitting}, where we show the 2D marginalized distributions in the $c_b$ vs.~$c_c$ and $c_g$ vs.~$c_c$ planes from a fit with flat priors for all parameters, allowing $|c_c|$ values as large as 5. We observe how the large prior for $c_c$ broadens the corresponding posteriors ---for the range displayed in the figure this is clearly visible in the 68\% probability regions---  a clear indication of overfitting~\cite{Wesolowski:2015fqa}. In the case of the $c_{Z\gamma}$, a flat prior is less problematic, because this coupling only enters in the $h\to Z\gamma$ decay but does not modify gluon fusion. We find an upper bound of around $35$ on its magnitude in this case. 
However, large values of the $|c_i|$ are clearly disfavoured by the EFT interpretation. If we are interested in observing the effects of natural deviations of $c_c$ from 1 without running into this kind of technical and interpretational issues one can do it so by, instead of fixing $c_c=1$, assuming a prior consistent with the EFT expected power counting. 

\begin{figure}[t]
   \begin{picture}(400,400)(0,0)
      \put(20,0){\includegraphics[width=400pt,trim=0 20 120 100,clip=true]{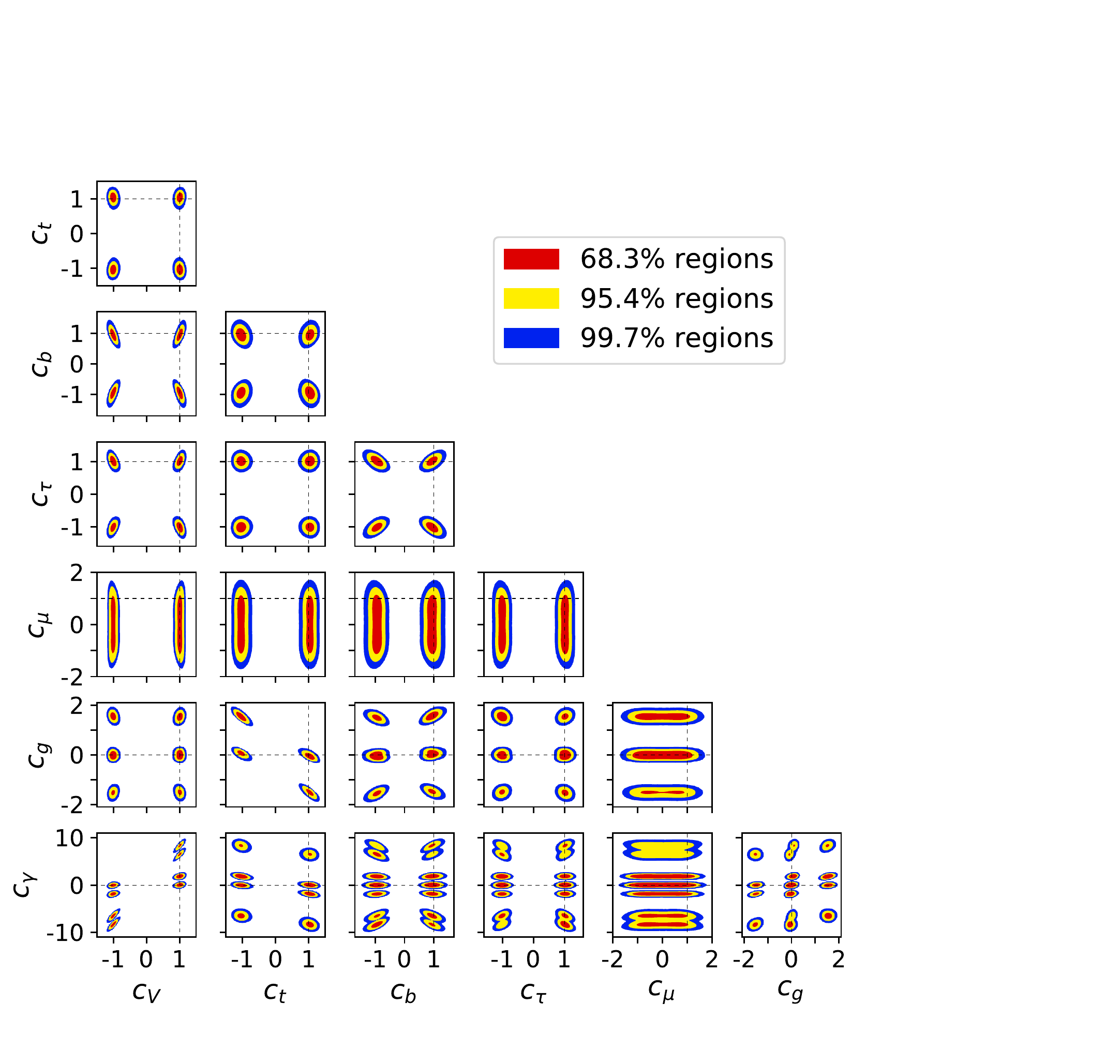}}
      \put(245,270){\includegraphics[width=50pt]{Figures/HEPfitLogo.png}}
   \end{picture}
   \caption{We show the 68.3\%, 95.4\% and 99.7\% probability regions in the $c_i$ vs.~$c_j$ planes for $i,j=V,t,b,\tau,\mu,g,\gamma$. We do not assume that the $c_i$ are close to the SM values, which are represented by the dashed black lines, but we use wide flat priors in order to find all possible regions compatible with the current $h$ signal strength bounds. Only for the experimentally poorly constrained $c_c$ and $c_{Z\gamma}$ we impose Gaussian priors with standard deviation $\sigma=0.5$. These choices correspond to the set of priors ${\cal P}_0$ described in eq.~(\ref{eq:P0}). See the text for details. 
   }
   \label{fig:Degeneracies}
\end{figure}

The results of such a fit are shown also in fig.~\ref{fig:Overfitting} and, for the rest of parameters, in fig.~\ref{fig:Degeneracies}. The posterior distributions and correlations in those figures were obtained assuming the following set of priors, denoted in what follows as ${\cal P}_0$:
\begin{equation}
{\cal P}_0:~~~
\begin{array}{l}
\mbox{Flat in}~c_{\psi,V} \in [-2,2],~~c_{g} \in [-3,3],~~c_{\gamma} \in [-12,12],\\
\\[-0.2cm]
\pi(c_{c,Z\gamma})\sim \exp{\left[-\frac {(c_{c,Z\gamma}-c_{c,Z\gamma}^{\SM})^2}{2~\! \sigma_{c, Z\gamma}^2}\right]},~\mbox{with}~\sigma_{c, Z\gamma}=0.5.
\end{array}
\label{eq:P0}
\end{equation}
The bounds on the flat priors in ${\cal P}_0$ are chosen to ensure that all regions allowed by the data are sampled, consistently with the small deviations in $c_c$, $c_{Z\gamma}$ permitted by their corresponding Gaussian priors. The exact choice of $\sigma_{c},~\!\sigma_{Z\gamma}=0.5$ as the standard deviations in the Gaussian priors in eq.~(\ref{eq:P0}) will be justified in the next section.
As expected, the result of the fit still shows multiple possible solutions.\footnote{In some cases the figure shows a slight difference in the probability content for some regions related by the invariances discussed in the text. This is just an 'artifact' of the MCMC sampling, common to instances where a problem has several disconnected solutions.} The approximate invariance $c_\psi \leftrightarrow -c_\psi$ is apparent for all fermionic couplings tested by the tree-level observables. All the fermionic couplings are consistent, in magnitude, with SM values. This includes also $c_\mu$, even though this coupling is still poorly constrained by data. The same applies to the coupling of the Higgs to the EW vector bosons. Also here, opposite sign couplings with respect to the SM are allowed by data. There are several possible patterns of (multi-parameter) correlations between those and non-SM values of the gluon and photon couplings $c_{g}$ ($c_\gamma$), allowing regions with values as large as $\pm2$ ($\pm10$). 
All such regions are, while consistent with experimental observation, unnatural from the point of view of the EFT used in the fit. In other words, the set of available observations and the precision of the employed data is not sufficient to constrain the \ewXL~parameters in a consistent way without extra information.


\subsection{Fits around the Standard Model solution}
\label{sec:SMsolution}

Now we concentrate on the region of the parameter space that is expected by the EFT power counting.
As a first step, we need to decide on how to treat the poorly constrained Wilson coefficients $c_c$ and $c_{Z\gamma}$.
From the previous section we know that too much freedom for $c_c$ leads to the problem that also the other parameters are ``artificially'' shifted away from the SM solution towards larger values in the fit.
Nevertheless, we want to allow for the possibility that $c_c$ and $c_{Z\gamma}$ can differ from their SM values.
The suppression of the mentioned overfitting regions can be achieved by assigning a Gaussian prior to both parameters~\cite{Jaynes:1957zza,Wesolowski:2015fqa}.
In order to find a reasonable value for the standard deviation of these priors, we choose Gaussian priors for all $c_i$ with a universal standard deviation $\sigma$. We therefore denote this set of normal prior distributions as ${\cal N}_\sigma$:
\begin{equation}
{\cal N}_{\sigma}:~~~
\pi(c_{i})\sim \exp{\left[-\frac {(c_{i}-c_{i}^{\SM})^2}{2~\! \sigma^2}\right]}~~~\forall i.
\end{equation}
This choice is also motivated by the naturalness argument: the $c_i$ are defined to be deviations from the SM limit, so one would expect them to have values around their SM values, in agreement with the \emph{principle of maximum entropy}~\cite{Jaynes:1957zza}.

\begin{figure}[t]
   \begin{picture}(400,400)(0,0)
      \put(0,0){\includegraphics[width=400pt]{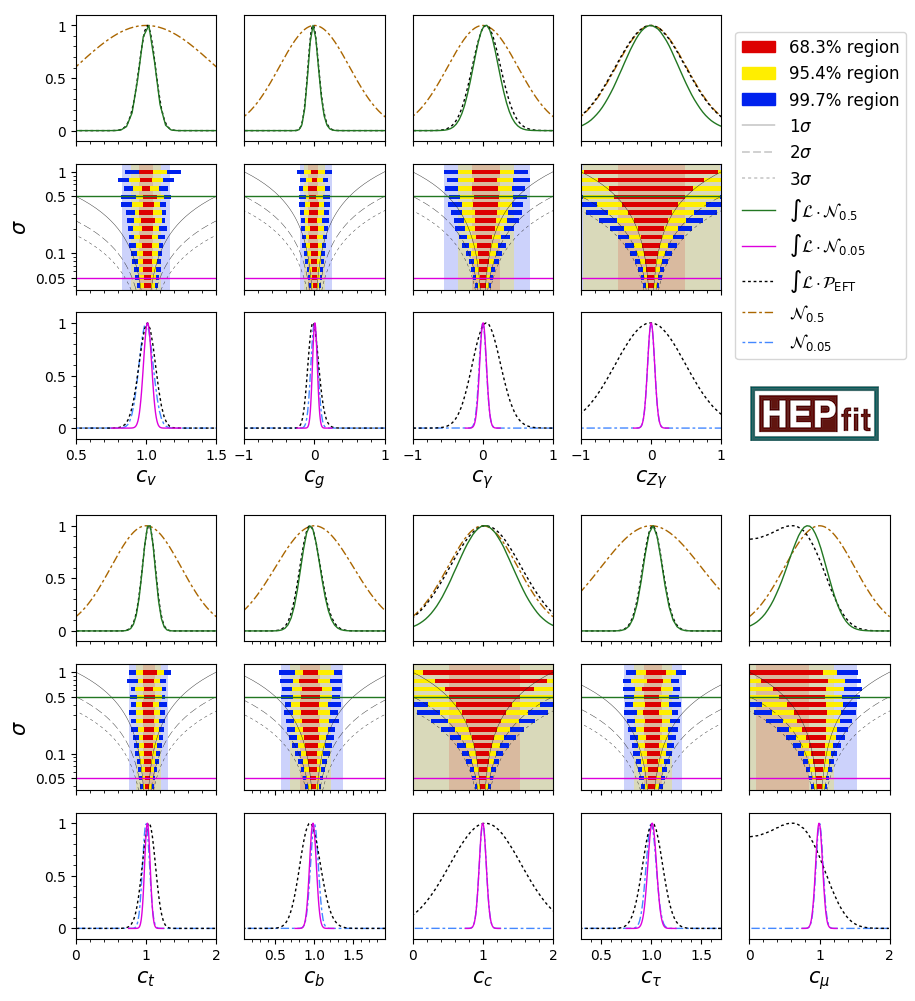}}
   \end{picture}
   \caption{Scan of the posterior for different universal widths of the Gaussian prior, inspired by \cite{Wesolowski:2015fqa}. For each coupling, we show how the posterior evolves with increasing width of the Gaussian priors in the central panel. The 68.3\%, 95.4\% and 99.7\% probability regions are red, yellow and blue bars, respectively; the transparent bands in the background represent the fit with the priors ${\cal P}_{\rm{EFT}}$. The upper (lower) panel of each coupling shows the ${\cal N}_{0.5}$ (${\cal N}_{0.05}$) prior in brown (light blue), its posterior in green (pink) and the ${\cal P}_{\rm{EFT}}$ posterior as the dotted black curve.}
   \label{fig:PriorDependence}
 \end{figure}

In fig.~\ref{fig:PriorDependence}, we show how the posterior distributions change~\cite{Wesolowski:2015fqa} if we vary the standard deviation of the priors ${\cal N}_\sigma$ from $10^{-1.4}$ to $10^{0}$.
For a better illustration, we explicitly highlight a case with a smaller $(10^{-1.3}\approx 0.05)$ and one with a larger $(10^{-0.3} \approx 0.5)$ standard deviation below and above each panel describing the dependence of the posteriors on $\sigma$.
The central value of the fit remains constant throughout the scan for all $c_{i}$ except for $c_{\mu}$ and $c_{c}$. 
For instance, the central value for $c_{\mu}$ moves closer to the SM for smaller $\sigma$, as the likelihood is highly asymmetric: values around $2$ are very unlikely, while values around $0$ are not excluded yet.

The most differences arise in the size of the error bars in the posterior. One can see that for $\sigma=0.05$ all posteriors simply reflect the priors and that all parameters, except for the charm and $Z\gamma$ couplings, become less prior dependent and more dominated by data as we move to larger values for the universal standard deviation.
However, for $\sigma > 0.5$, the upper 68\% probability limit of the $c_c$ posterior exceeds the upper 68\% probability limit of its prior. Also, the posterior distributions of $c_V$ and $c_t$ tend to have a larger upper limit. Both effects signal the presence of overfitting.
We therefore decided that $\sigma =0.5$ is a good compromise which allows as much freedom as possible for $c_c$ and $c_{Z\gamma}$ but at the same time keeps the overfitting error sufficiently small.
Since we want to keep the prior dependence of our fit as small as possible, we refrain from using the universal Gaussian priors for all other parameters ($c_V$, $c_t$, $c_b$, $c_\tau$, $c_\mu$, $c_g$ and $c_\gamma$) in the next step.
We only decrease the range of the flat priors from the previous section and allow for a maximal deviation of 1 from the SM value.
Furthermore, we assume that $c_{t}+c_{g}> 0$ in order to remove a second, non-SM-like solution in the $c_g$ vs.~$c_t$ plane (see fig.~\ref{fig:Degeneracies}).
With all these, we eliminate all but the SM-like solutions in the fit. We denote this choice of prior as ${\cal P}_{\rm{EFT}}$ in the following,
\begin{equation}
{\cal P}_{\rm{EFT}}:~~~
\begin{array}{l}
\mbox{Flat in}~c_{i}=c_{i}^\SM \pm 1~~~\text{for } i = V, t, b, \tau, \mu, g, \gamma,\\
\\[-0.4cm]
c_t+c_g>0,\\
\pi(c_{c,Z\gamma})\sim \exp{\left[-\frac {(c_{c,Z\gamma}-c_{c,Z\gamma}^{\SM})^2}{2~\! \sigma_{c, Z\gamma}^2}\right]},~\mbox{with}~\sigma_{c, Z\gamma}=0.5.
\end{array}
\label{eq:PEFT}
\end{equation}

The result of a fit with these priors 
can be found in the background of the sigma dependent panels in fig.~\ref{fig:PriorDependence} and as black dotted line in the posterior distribution planes.
Comparing it to the fit with universal Gaussian priors with $\sigma=0.5$ (upper panels), we observe that both fits are in good agreement for most parameters.
The overfitting error is visible in the corresponding $c_c$ plane, where the ${\cal P}_{\rm{EFT}}$ posterior slightly deviates from the ${\cal N}_{0.5}$ prior, preferring larger values for $c_c$.
The muon coupling $c_\mu$ is the parameter which deviates most from its SM value 1.
Here, the discrepancy between the ${\cal N}_{0.5}$ and the ${\cal P}_{\rm{EFT}}$ posterior is the largest and qualifies the choice of flat instead of Gaussian priors for all parameters except for $c_c$ and $c_{Z\gamma}$. In any case, with the exception of these parameters, the results of the fit are {\em prior independent}.

\begin{table}[t!]
\begin{center}
\begin{tabular}{ c | c c c }
\toprule
Parameter & \multicolumn{3}{c}{Fit results}\\
 &Full dataset& Run 1 only & Run 2 only\\
\midrule
$c_V$ & $1.01\pm 0.06$ & $0.96^{+0.08}_{-0.09}$ & $1.04\pm 0.08$ \\
$c_t$& $1.04^{+0.09}_{-0.10}$ & $1.16^{+0.32}_{-0.43}$ & $1.04\pm 0.10$ \\
$c_b$ & $0.95\pm 0.13$ & $0.93^{+0.24}_{-0.22}$ & $1.02^{+0.18}_{-0.17}$ \\
\crowcolor$c_c$ & $1.04\pm 0.51$ & $1.04\pm 0.51$ & $1.04\pm 0.51$ \\
$c_\tau$ & $1.02\pm 0.10$ & $1.04^{+0.15}_{-0.14}$ & $1.02\pm 0.15$ \\
$c_\mu$ & $0.58^{+0.40}_{-0.38}$ & $<1.1$ @ 68\% & $0.60^{+0.42}_{-0.40}$ \\
&&($<1.76$ @ 95\%)&\\
$c_g$ & $-0.01^{+0.08}_{-0.07}$ & $-0.12^{+0.35}_{-0.27}$ & $0.02^{+0.09}_{-0.08}$ \\
$c_\gamma$ & $0.05\pm 0.20$ & $-0.34^{+0.47}_{-0.38}$ & $0.22^{+0.25}_{-0.24}$ \\
\crowcolor$c_{Z\gamma}$ & $-0.01\pm 0.50$ & $0.00\pm 0.50$ & $-0.01\pm 0.50$ \\
\bottomrule
\end{tabular}
\caption{Results from the fits around the SM solution using the set of priors denoted as ${\cal P}_{\rm{EFT}}$. 
These are flat priors in the range  $c_i=c_i^{\mathrm{SM}} \pm 1$ and $c_t+c_g>0$ for all parameters except for $c_c$ and $c_{Z\gamma}$, which have Gaussian priors with $\sigma=0.5$, see eq.~(\ref{eq:PEFT}).
The results for $c_c$ and $c_{Z\gamma}$ (identified by shades of gray in the table) are fully prior dominated.
For each parameter we quote the median and error ---defined from the boundaries of the 68\% probability interval--- computed from the corresponding 1D posterior distributions.
}
\label{tab:fitresults}
\end{center}
\end{table}

\begin{table}[t!]
\begin{tabular}{c | c c c c c c c c c}
\toprule
&      $c_V$&$c_t$&$c_b$&$c_c$&$c_\tau$&$c_\mu$&$c_g$&$c_\gamma$&$c_{Z\gamma}$\\
\midrule
$c_V$&1~~&~~ 0.12~~&~~ 0.71~~&~~ 0.25~~&~~ 0.49~~&~~ 0.09~~&~~ 0.14~~&~~ 0.32~~&~~ 0\\
$c_t$&0.12~~&~~ 1~~&~~ 0.25~~&~~ 0.16~~&~~ 0.05~~&~~ 0.01~~&~~ -0.68~~&~~ -0.31~~&~~ 0\\
$c_b$&0.71~~&~~ 0.25~~&~~ 1~~&~~ 0.09~~&~~ 0.56~~&~~ 0.07~~&~~ 0.36~~&~~ 0.03~~&~~ 0\\
$c_c$&0.25~~&~~ 0.16~~&~~ 0.09~~&~~ 1~~&~~ 0.14~~&~~ 0.02~~&~~ 0.04~~&~~ 0.03~~&~~ 0\\
$c_\tau$&0.49~~&~~ 0.05~~&~~ 0.56~~&~~ 0.14~~&~~ 1~~&~~ 0.06~~&~~ 0.25~~&~~ 0.01~~&~~ 0\\
$c_\mu$&0.09~~&~~ 0.01~~&~~ 0.07~~&~~ 0.02~~&~~ 0.06~~&~~ 1~~&~~ 0~~&~~ 0~~&~~ 0\\
$c_g$&0.14~~&~~ -0.68~~&~~ 0.36~~&~~ 0.04~~&~~ 0.25~~&~~ 0~~&~~ 1~~&~~ 0.33~~&~~ 0\\
$c_\gamma$&0.32~~&~~ -0.31~~&~~ 0.03~~&~~ 0.03~~&~~ 0.01~~&~~ 0~~&~~ 0.33~~&~~ 1~~&~~ 0\\
$c_{Z\gamma}$&0~~&~~ 0~~&~~ 0~~&~~ 0~~&~~ 0~~&~~ 0~~&~~ 0~~&~~ 0~~&~~ 1\\
\bottomrule
\end{tabular}
\caption{Correlation matrix for the fit around the SM solution using the set of priors denoted as ${\cal P}_{\rm{EFT}}$ and the full data set of Higgs observables from Tevatron and the LHC runs 1 and 2 (first column in table~\ref{tab:fitresults}. 
}
\label{tab:CorrC}
\end{table}

After finding the most suitable choice for the priors, we want to discuss in detail the resulting simultaneous fit to all $c_i$.
From the posterior of the fit we compute the median for all parameters, as well as their 68\% probability uncertainties. For each $c_i$, the latter are defined from the upper and lower boundaries of the 68\% probability interval, after marginalizing over the other parameters. 
The numerical values of the results of the fit can be found in the second column of table \ref{tab:fitresults}.
For a comparison, we also list the results for individual fits to only run 1 data and only run 2 data in the third and fourth column.\footnote{What we label run 1 here and in the following also contains two Tevatron analyses~\cite{Aaltonen:2013ipa,Abazov:2013gmz}.}
We observe that in the combined fit, the coupling to vector bosons $c_V$ is now determined with a precision of 6\%.
The uncertainties of the Higgs couplings to third generation fermions and gluons are around 10\%.
For the measurement of the Higgs coupling to coloured particles, the bounds from run 2 data are stronger.
While after run 1 one could only extract an upper limit on $|c_\mu|$, the run 2 constraints allow us to determine the muon Wilson coefficient with a precision of 40\%. (As already mentioned, its central value is the only which visibly deviates from SM expectations, but not at a significant level.)
For the photon coupling, we observe that the results using run 1 and run 2 data, individually, feature small deviations from the SM limit.
The results from the two data sets are, however, also in slight tension with each other, with run 1 (run 2) data preferring $c_\gamma <c_\gamma^\SM$ ($> c_\gamma^\SM$).
In the combined fit, both preferences average to a central value close to the SM, with an uncertainty of $\pm 0.20$.
Finally, for the $c_c$ and $c_{Z\gamma}$ posteriors we get essentially the prior distributions; only the central value of the charm coupling is shifted by 4\% towards larger values, being an effect of the above-mentioned tendency to overfitting.
If we use $\sigma_c=1$ instead of $\sigma_c=0.5$ for $c_c$, the shift of the central value amounts to 16\%.

\begin{figure}[t]
   \begin{picture}(400,400)(0,0)
      \put(0,0){\includegraphics[width=435pt]{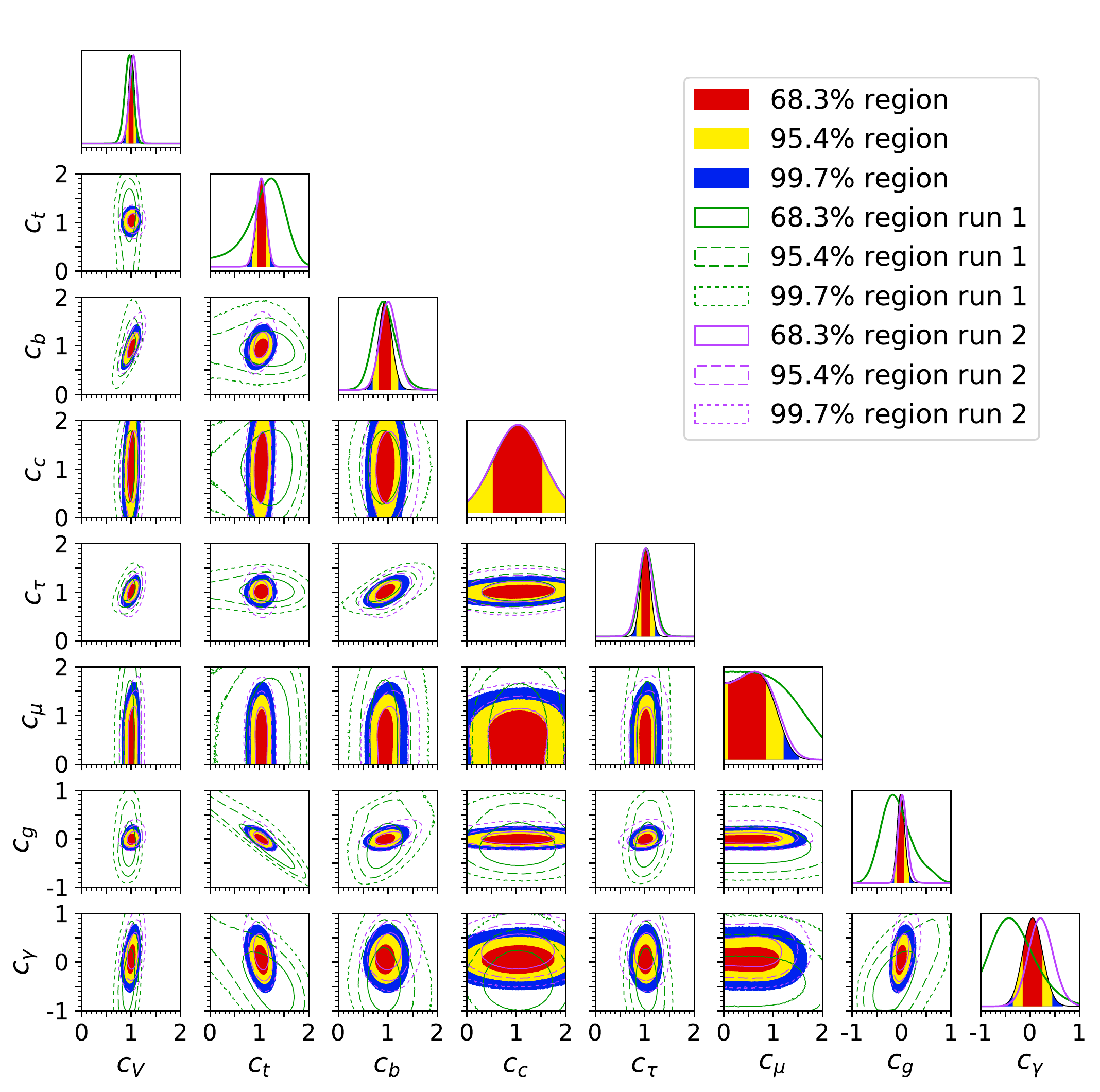}}
      \put(271,230){\includegraphics[width=50pt]{Figures/HEPfitLogo.png}}
   \end{picture}
   \caption{For the parameters $c_i$ with $i=V,t,b,c,\tau,\mu,g,\gamma$ we display the one-dimensional posterior distribution as well as their two-dimensional correlations. The regions allowed at 68.3\%, 95.4\% and 99.7\% probability by current Higgs data are represented by the red, yellow and blue filled contours, respectively. Additionally, we show the single contributions from pre-13 TeV run data (green) and LHC run 2 data (purple).}
   \label{fig:Triangle}
 \end{figure}

Apart from the median values and the 68\% probability allowed ranges we also want to address the two-dimensional correlations between all parameters. 
These numerical correlations are given in table~\ref{tab:CorrC} and illustrated in fig.~\ref{fig:Triangle}.
The matrix in fig.~\ref{fig:Triangle} also contains the information about the one-dimensional posterior distributions of our fit to run 1, run 2 and the combined data sets on the diagonal as a graphical translation of table~\ref{tab:fitresults}.
The off-diagonal panels depict the corresponding two-dimensional posterior distributions in the $c_i$ vs.~$c_j$ planes if we marginalize over all other parameters.
As described above, the result for $c_{Z\gamma}$ it is completely dominated by its Gaussian prior. Also, from table~\ref{tab:CorrC} we see there are no noticeable correlations with this parameter in the results. Because of that, we do not include this parameter in fig.~\ref{fig:Triangle}.

Addressing correlations between the parameters, we can see that, due to the interplay in the gluon fusion production mechanism, the top and the gluon coupling are strongly anti-correlated.
Also due to the loop couplings, but to a lesser extent, $c_\gamma$ is anti-correlated with $c_t$ and therefore correlated with $c_g$.
Further correlations worth mentioning can be found between $c_V$, $c_b$ and $c_\tau$.
Comparing the fits to only run 1 and only run 2 data, we find that all three $c_t$ contours are fully contained in the chosen range for the latter, while $c_t=0$ was allowed at 95\% probability by run 1 signal strengths only.
The reason for this is the improvement in the experimental sensitivity to tth production. In particular, the latest ATLAS results show evidence of this mechanism consistent with the SM~\cite{Aaboud:2017jvq}, which helps to resolve the flat direction in the $c_g$ vs.~$c_t$ plane. 
Therefore, this is also correlated with the reduction of the $c_g$ uncertainties. 
The parameter $c_\mu$ must be smaller than 2 at 99.7\% probability according to the run 2 measurements, while such a large value was compatible at the 95\% probability level with run 1 data.
Finally, as mentioned above, we can clearly see that run 1 and run 2 pull $c_\gamma$ into opposite directions, whereas the combined contours are centered around the SM limit and well within the range between $-1$ and $+1$.

As a cross check, we compared our run 1 results with those existing in the literature and, in particular, with previous independent work from one of the authors in Ref.~\cite{Buchalla:2015qju}, where a fit to run 1 data was discussed in the context of the electroweak chiral Lagrangian. Comparing the results presented there with our run 1 fit, we see some small differences. These are, however, understood from an improved treatment of the tth production channel and differences in the treatment of the experimental information used as inputs in the fits: Ref.~\cite{Buchalla:2015qju} used the fitted signal strengths per production modes, while here we use directly the experimental information from all categories that enter in such fits. At any rate, all values found in~\cite{Buchalla:2015qju} are within the $68\%$ probability region of our fit.


\section{Projections for future Higgs factories}
\label{sec:projections}

After having scrutinized the current limits on the Higgs electroweak chiral Lagrangian parameters $c_i$ in the previous section, we would like to address here the potential for improvements of such constraints at the end of the LHC life-cycle (see \cite{Englert:2015hrx,Englert:2017aqb} for studies including also the information from differential Higgs distributions) and at future Higgs factories. We therefore consider several scenarios: 
\begin{itemize}
{\item The High-Luminosity upgrade of the LHC (HL-LHC), where the precision for those Higgs observables whose uncertainty is statistically dominated could be largely improved. We use the precisions for the signal strength modifiers corresponding to an integrated luminosity of 3 ab$^{-1}$ in Refs.~\cite{ATL-PHYS-PUB-2013-014,ATL-PHYS-PUB-2014-011,ATL-PHYS-PUB-2014-016,CMS-NOTE-2013-002}.}
{\item The International Linear Collider (ILC) is designed as an $e^+ e^-$ Higgs factory. The current operation baseline assumes collisions at $\sqrt{s}=250$ GeV. The possibility of running at 500 GeV and 1 TeV centre-of-mass energies has been also discussed, as well as the possibility of a luminosity upgrade, which we consider here. We use as ILC inputs the precisions detailed in \cite{Dawson:2013bba}. We label the 250 GeV scenario, assuming full luminosity (1.15 ab$^{-1}$), as ``ILC-250''. We also consider an scenario corresponding to using the data accumulated during all the three stages discussed in \cite{Dawson:2013bba}: a total 5.25 ab$^{-1}$ of data, distributed in 1.15 ab$^{-1}$ at $\sqrt{s}=250$ GeV, 1.6 ab$^{-1}$ at 500 GeV and 2.5 ab$^{-1}$ at 1 TeV. We denote this scenario as ``ILC-all''. }
{\item The Future Circular Collider project  at CERN, and in particular the $e^+e^-$ collider option (FCC-ee). The projected work points include centre-of-mass energies of $\sqrt{s}=240$ GeV and 350 GeV, where we could be sensitive to $e^+ e^- \to Zh$ and $e^+ e^- \to \nu\bar{\nu}h$ production. The FCC-ee envisions the largest luminosity of all projected future $e^+ e^-$ machines (10 ab$^{-1}$ of data at 240 GeV and 2.6 ab$^{-1}$ at 350 GeV, assuming 4 interaction points). The projected experimental precisions have been extracted from \cite{Gomez-Ceballos:2013zzn}.}
{\item The Circular Electron Positron Collider (CEPC) is another $e^+e^-$ circular collider which would be based in China. Like the FCC-ee, it contemplates a ``Higgs-factory'' run at $\sqrt{s}=250$ GeV, but there is no information about a 350 GeV run in the current project design~\cite{CEPC-SPPCStudyGroup:2015csa}.
The total accumulated luminosity at 250 GeV is expected to be of the order of 5 ab$^{-1}$. All the precisions for the Higgs observables are taken from~\cite{CEPC-SPPCStudyGroup:2015csa}.}
{\item Finally, we also consider the proposed design of the Compact Linear Collider (CLIC) at CERN. This is also an $e^+ e^-$ linear collider, with a particular focus on the exploration at high energies. The different CLIC runs would operate at $\sqrt{s}=$380 GeV, 1.4 TeV and 3 TeV. As in the ILC case, we choose one scenario based on 0.5 ab$^{-1}$ of data from the lowest energy run at 380 GeV (``CLIC-380'')\footnote{The actual numbers presented in Ref.~\cite{Dawson:2013bba,Abramowicz:2016zbo} were obtained assuming a center of mass energy of 350 GeV.}, and one scenario 
adding also the 1.5 ab$^{-1}$ and 2 ab$^{-1}$ of data taken at 1.4 TeV and 3 TeV, respectively (``CLIC-all'')~\cite{Dawson:2013bba,Abramowicz:2016zbo}.}
\end{itemize}
For all future lepton machines it is expected that, apart from reconstructing the different decay channels, one could also measure the total $e^+ e^- \to Zh$ cross section using the distribution of the mass recoiling against the $Z$. This is the dominant production channel around $250$ GeV. At 350 GeV, while still small compared to $Zh$, the cross section of $e^+ e^- \to \nu\bar{\nu}h$ production via $W$ boson fusion is already sizable to provide sensitivity to this mode. The energies attainable at future circular $e^+ e^-$ colliders are however, not enough to open the $tth$ production mode, and hence direct sensitivity to modifications of the top Yukawa coupling. This is only possible at linear colliders. 
In addition to an increased precision on the couplings, the future colliders data sets will also start to constrain further parameters that we neglected in our analysis because they are currently not accessible. An example is the Higgs self-coupling, see \cite{DiVita:2017vrr,Kim:2018uty}. As we are only interested in this study in the comparison of present and future bounds on Higgs couplings, and in this regard double-Higgs production would not add a significant improvement, we also ignore such measurements in our study of the precisions at future colliders.

\begin{table}[t!]
\begin{center}
\begin{tabular}{ c | c | c | c c c c c c}
\toprule
Collider 
 & LHC now & \cellcolor{color1}HL-LHC & \cellcolor{color2}ILC & \cellcolor{color2}ILC & \cellcolor{color3}CLIC & \cellcolor{color3}CLIC & \cellcolor{color4}CEPC & \cellcolor{color5}FCC-ee \\[-2pt]
 & & \cellcolor{color1} & \cellcolor{color2}250 & \cellcolor{color2}all & \cellcolor{color3}380 & \cellcolor{color3}all & \cellcolor{color4} & \cellcolor{color5} \\[-2pt]
$L$ [ab$^{-1}$] & 0.06 & \cellcolor{color1}3 & \cellcolor{color2}1.2 & \cellcolor{color2}5.3 & \cellcolor{color3}0.5 & \cellcolor{color3}4 & \cellcolor{color4}5 & \cellcolor{color5}12.6 \\
\midrule
$c_V$ & $60$ & $30$ & $3.0$ & $1.2$ & $4.4$ & $1.6$ & $1.6$ & $1.1$ \\
&&& ($3.5$) & ($1.2$) & ($5.6$) & ($1.8$) & ($1.7$) & ($1.1$) \\[3pt]
\hline
$c_t$& $100$ & $53$ & $52$ & $17$ & $53$ & $32$ & $52$ & $51$ \\
&&& (--) & ($19$) & (--) & ($40$) & (--) & (--) \\[3pt]
\hline
$c_b$ & $130$ & $39$ & $8.5$ & $3.1$ & $11$ & $3.0$ & $5.0$ & $3.3$ \\
&&& ($11$) & ($3.2$) & ($19$) & ($3.2$) & ($5.5$) & ($3.5$) \\[3pt]
\hline
$c_c$ & -- & -- & $21$ & $8.4$ & $63$ & $22$ & $12$ & $6.9$ \\
&&& ($23$) & ($8.5$) & ($67$) & ($22$) & ($13$) & ($7.0$) \\[3pt]
\hline
$c_\tau$ & $100$ & $40$ & $12$ & $6.9$ & $22$ & $13$ & $7.7$ & $4.8$ \\
&&& ($15$) & ($7.2$) & ($38$) & ($15$) & ($8.2$) & ($4.9$) \\[3pt]
\hline
$c_\mu$ & $400$ & $62$ & $53$ & $46$ & $53$ & $47$ & $45$ & $41$ \\
&&& (--) & ($100$) & (--) & ($110$) & ($88$) & ($66$) \\[3pt]
\hline
$c_g$ & $80$ & $41$ & $40$ & $14$ & $40$ & $25$ & $40$ & $39$ \\
&&& (--) & ($15$) & (--) & ($32$) & (--) & (--) \\[3pt]
\hline
$c_\gamma$ & $200$ & $75$ & $63$ & $41$ & $66$ & $49$ & $59$ & $54$ \\
&&& (--) & ($79$) & (--) & ($140$) & (--) & (--) \\[3pt]
\hline
$c_{Z\gamma}$ & -- & $950$ & $900$ & $530$ & $900$ & $760$ & $920$ & $920$ \\
&&& (--) & (--) & (--) & ($1000$) & (--) & (--) \\
\midrule
$ $&\multicolumn{8}{c}{Sensitivity/uncertainty [$\times 10^{-3}$]}\\
\bottomrule
\end{tabular}
\caption{Projections for the 68\% probability sensitivity to deviations from the SM Higgs interactions at various future colliders. The ILC, CLIC, CEPC and FCC numbers also include the final HL-LHC results; their individual contributions are given in parentheses. We also show, for comparison, the current LHC result. (All uncertainties on the EFT coefficients are in units of $10^{-3}$.)
}
\label{tab:futureresults}
\end{center}
\end{table}

\begin{figure}[t]
   \begin{picture}(400,400)(0,0)
      \put(0,0){\includegraphics[width=435pt]{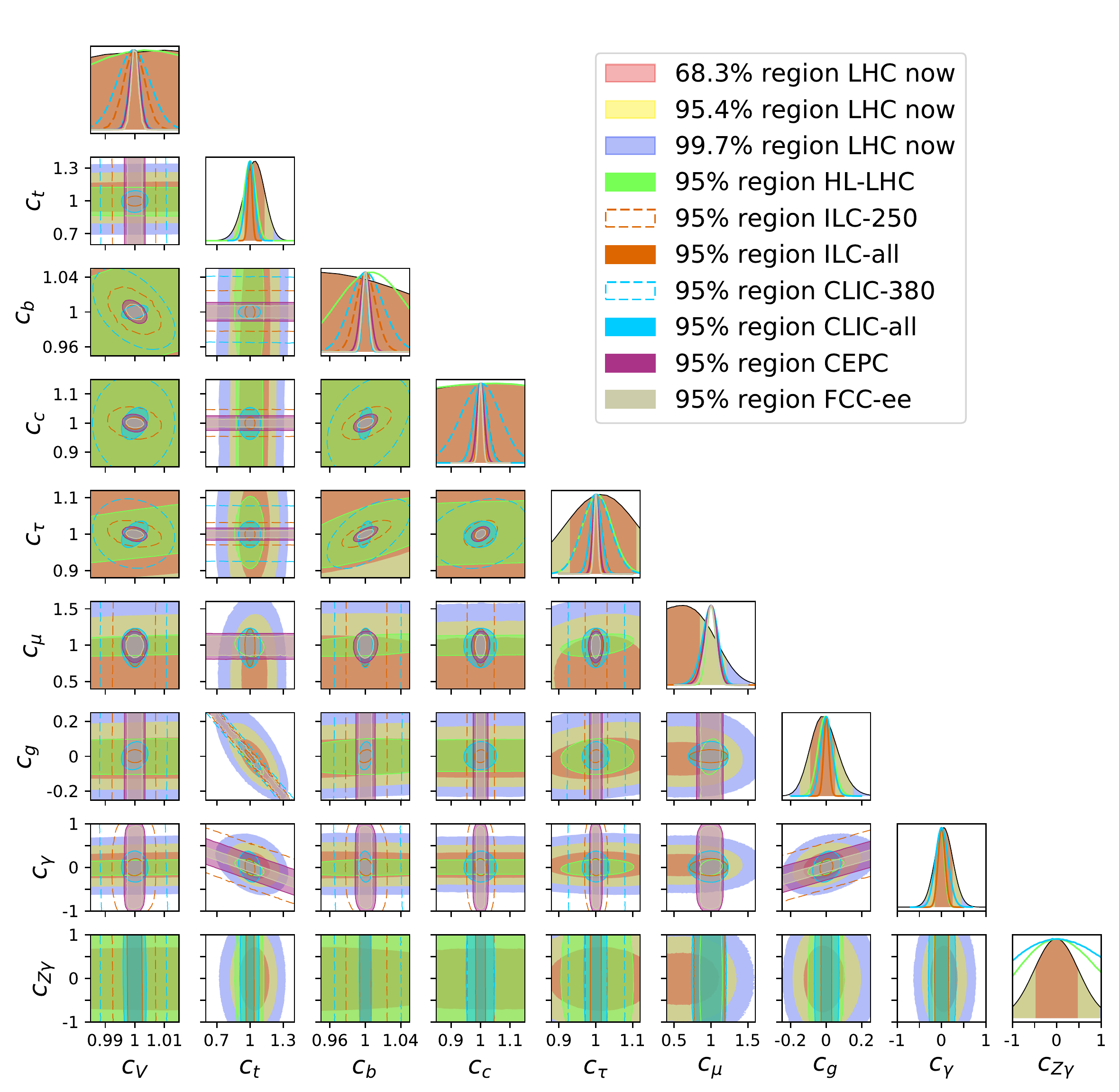}}
      \put(295,200){\includegraphics[width=50pt]{Figures/HEPfitLogo.png}}
   \end{picture}
   \caption{On top of the fit to current data in the background (same as in fig.~\ref{fig:Triangle}) we illustrate the presumed impact of future colliders on the parameters $c_i$ with $i=V,t,b,c,\tau,\mu,g,\gamma,Z\gamma$. For the future projections, we only show the $95$\% probability regions; for the corresponding colours we refer to the legend and to table \ref{tab:futureresults}. (Note that, while current results for $c_c$ and $c_{Z\gamma}$ are obtained using the priors ${\cal P}_\mathrm{EFT}$, we use flat priors for the calculation of future uncertainties.)
  }
   \label{fig:Future}
\end{figure}

In the fits presented in this section we use flat priors for all $c_i$, and we construct the likelihood assuming Gaussian distributions around the SM values for the future signal strength measurements, with errors given by the corresponding future experimental uncertainty.
The experimental inputs for these uncertainties, as obtained from the corresponding references given above, are collected in appendix~\ref{app:futcol_data}. 
The numerical results for the sensitivities, defined as the 68$\%$ probability uncertainty on the fit parameters, are given in table \ref{tab:futureresults}.
The results for the future lepton colliders are presented combined with the HL-LHC projections. To illustrate the individual constraining capabilities of each type of collider, we also show in parentheses the results obtained without the HL-LHC information.\footnote{Since the projections for the uncertainties at future lepton colliders are essentially dominated by statistics, the basic results presented here can be scaled in a straightforward manner to scenarios with different luminosity assumptions (or, for circular collider, also different number of interaction points).}
The corresponding one-dimensional and two-dimensional posterior distributions from each individual collider can be found in fig.~\ref{fig:Future}, in which we also show the current distributions from fig.~\ref{fig:Triangle} in the background.
For the purpose of comparing the different collider options it is very important to keep in mind that the ILC results were derived {\it assuming} a high-luminosity upgrade, as detailed in \cite{Dawson:2013bba}, while for the other colliders only information about the baseline options is available. On a related note, for the comparison between the FCC-ee and CEPC results one must also take into account that the precision on the Higgs observables for the former in~\cite{Gomez-Ceballos:2013zzn} assume 4 interaction points, compared to only 2 for CEPC. Note however that, even rescaling the luminosities to 2 interaction points ---thus equating the luminosities at 240 GeV to 5 ab$^{-1}$ for both circular colliders--- the FCC-ee has an advantage due to the extra measurements that would be taken during the 350 GeV run. 

From our results one can see that the coefficients $c_V$, $c_b$, $c_c$ and $c_\tau$ could be measured to a high precision at both linear and circular electron positron colliders. The latter cannot disentangle the correlations of $c_t$, $c_g$ and $c_\gamma$, for which the strongest future bounds would come from the HL-LHC or linear colliders. What can also be extracted from the fits are upper limits on certain combinations of these three parameters. The strongest bound here is the one on $c_t+c_g$, which can be determined with a precision of $\sim 1\%$ at linear colliders.
All future colliders provide limits on $c_\mu$ of the same order. Finally, weak bounds on $c_{Z\gamma}$ can be extracted from HL-LHC data and, with even lower precision, from all combined CLIC measurements. While the other machines could in principle also observe the $Z\gamma$ decay mode, there is no official information to asses their sensitivity to the corresponding coefficient. 


\section{Application to minimal composite Higgs models}
\label{sec:CHM}

A popular solution to the hierarchy problem are composite Higgs models (CHM). The Higgs emerges as pseudo-Nambu-Goldstone boson of a global symmetry breaking at the scale $f\geq v$ in these scenarios. Because of their strong dynamics, CHMs are best described by the electroweak chiral Lagrangian at low energies. In this section we therefore use the results previously obtained in term of the general \ewXL~parameterization to estimate the constraints on some of these composite Higgs scenarios.~\footnote{For a recent dedicated analysis of the Higgs signal strengths within the context of minimal composite Higgs models see also \cite{Banerjee:2017wmg}.}

\begin{figure}[t]
   \begin{picture}(300,300)(0,0)
      \put(60,0){\includegraphics[width=300pt]{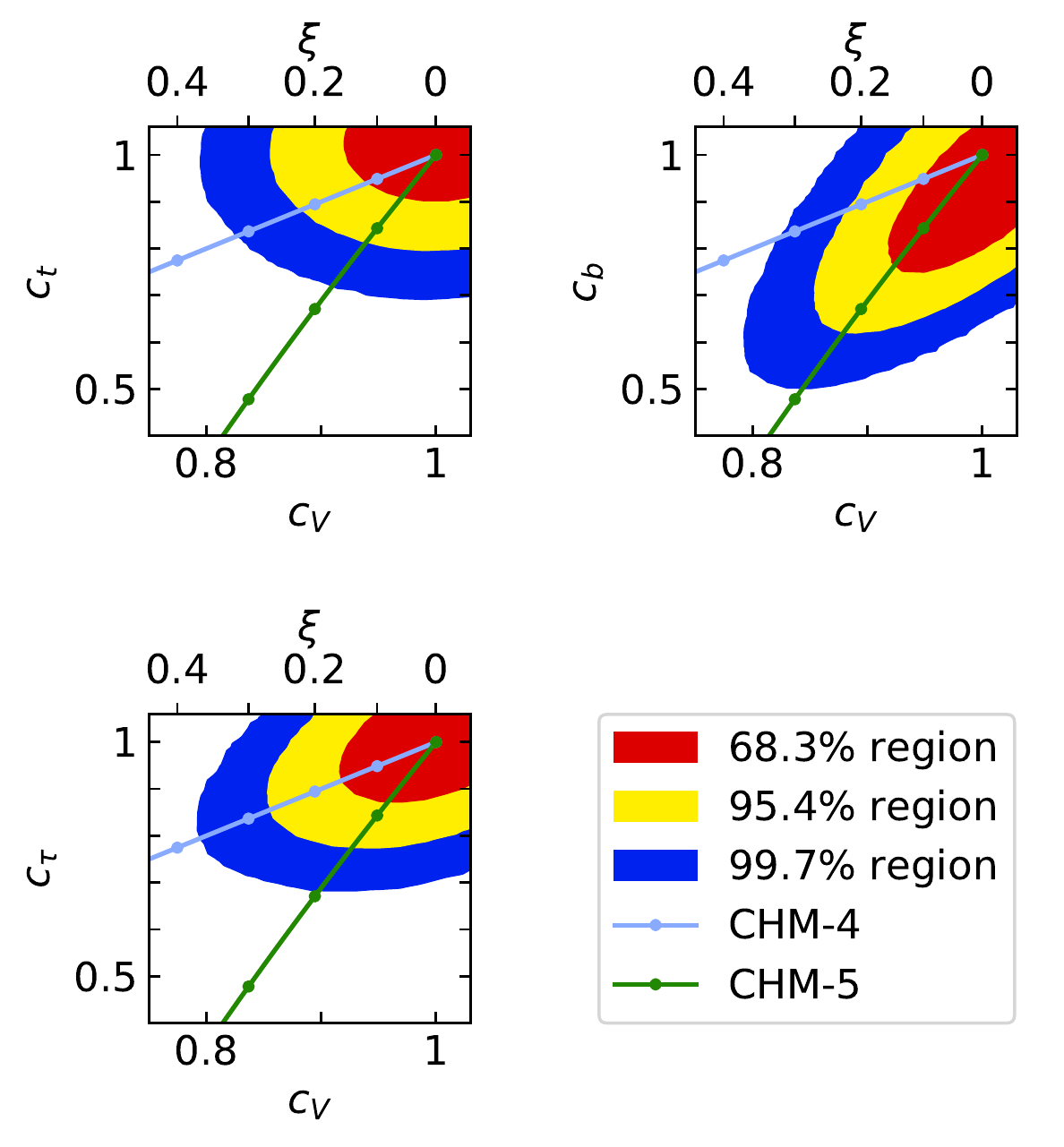}}
      \put(260,130){\includegraphics[width=50pt]{Figures/HEPfitLogo.png}}
   \end{picture}
   \caption{Magnification of the $c_\psi$ vs.~$c_V$ planes from fig.~\ref{fig:Triangle}, together with the lines corresponding to the two composite Higgs scenarios discussed in the text. The dots on the lines correspond to values of $\xi=$0, 0.1, 0.2, 0.3, 0.4 in the direction from $c_V=1$ downwards.}
   \label{fig:MCHMnow}
 \end{figure}

\begin{figure}[t]
   \begin{picture}(400,300)(0,0)
      \put(0,0){\includegraphics[width=400pt]{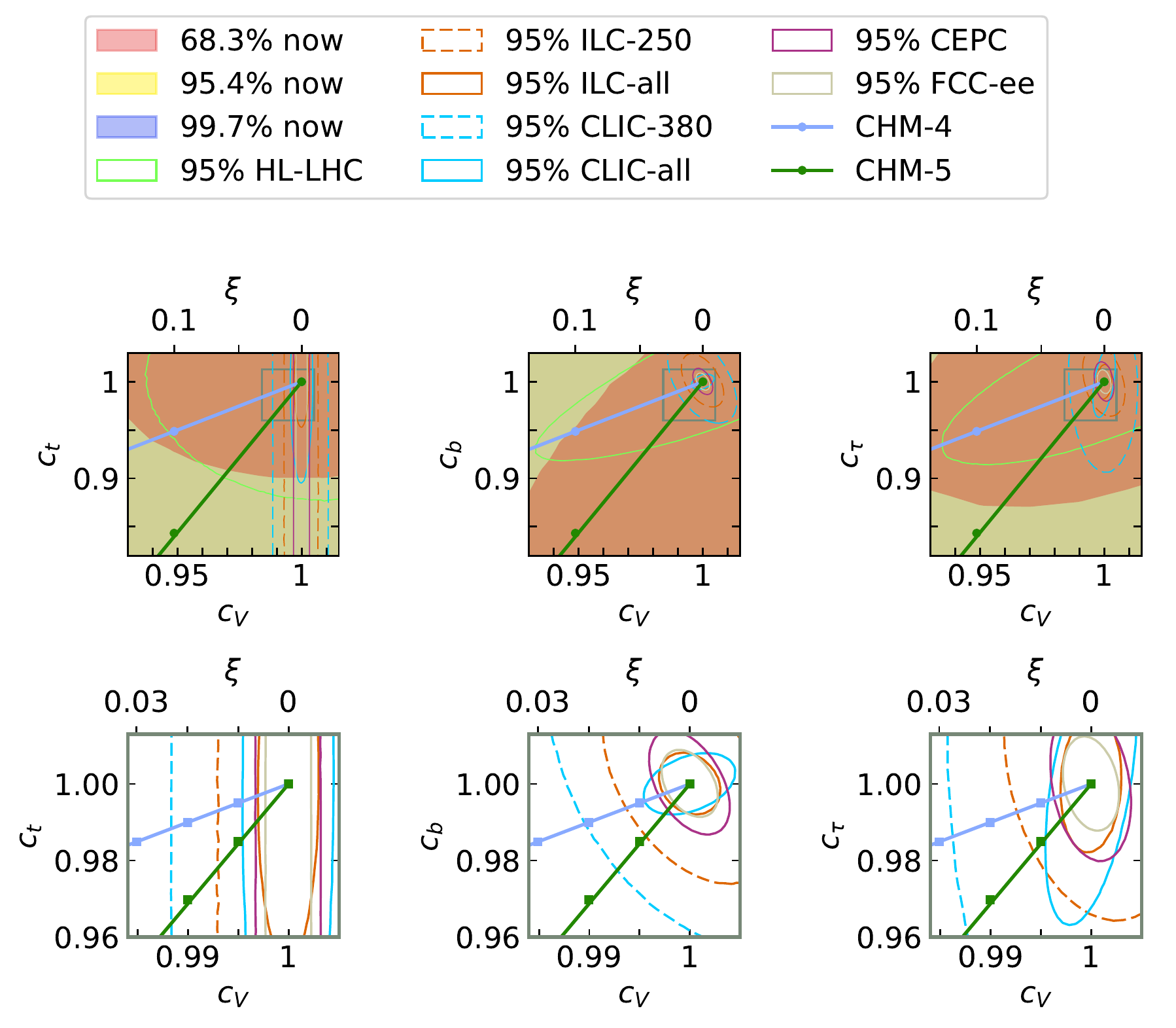}}
      \put(120,257){\includegraphics[width=50pt]{Figures/HEPfitLogo.png}}
   \end{picture}
   \caption{Magnification of the $c_\psi$ vs.~$c_V$ planes from fig.~\ref{fig:Future}, together with the lines corresponding to the two composite Higgs scenarios discussed in the text. In the top row, we show enlarged regions for the third generation fermion couplings, in which the dots are equivalent to the ones in fig.~\ref{fig:MCHMnow} for $\xi=$0, 0.1 in decreasing $c_V$ direction. The bottom row zooms in on the grey boxes of the top row. Here the squares mark the points $\xi=$0, 0.01, 0.02, 0.03.}
   \label{fig:MCHMfuture}
 \end{figure}

The so-called minimal composite Higgs models are based on the coset $SO(5)/SO(4)$~\cite{Agashe:2004rs,Contino:2006qr,Contino:2010rs,Carena:2014ria}\footnote{See \cite{Buchalla:2014eca} for an operator matching to the \ewXL.}.  The coupling of the Higgs to the weak gauge bosons arises from the kinetic term of the Goldstone bosons and, in this minimal $SO(5)/SO(4)$-scenario, it has the form  
\begin{equation}
c_{V}=\sqrt{1-\xi},~\mbox{with}~\xi = v^{2}/f^{2}. 
\end{equation}
The couplings of the Higgs to the fermions depend on the $SO(5)$ representation where the SM fermions are embedded, and are therefore model-dependent. The smallest representations are the $\mathbf{4}$ \cite{Agashe:2004rs} and the $\mathbf{5}$ \cite{Contino:2006qr}. In these, the fermion-Higgs coupling becomes~\cite{Azatov:2012bz}
\begin{equation}
\label{eq:CHM}
c_{\psi}^{(4)}=\sqrt{1-\xi}\qquad  \text{and}\qquad c_{\psi}^{(5)}= \frac{1-2\xi}{\sqrt{1-\xi}},
\end{equation}
respectively. (Other cosets and representations may also exhibit a similar structure, see \cite{Pomarol:2012qf,Sanz:2017tco} for other examples and generalizations.) Since in these two cases the couplings $c_{V}$ and $c_{\psi}$ depend only on the parameter $\xi$, the parameter space of these models corresponds to a line in the $c_{\psi}$ vs.~$c_{V}$ plane. We show them labeled as CHM-4 (CHM-5) for third-generation fermions in the $\mathbf{4}$ ($\mathbf{5}$) representation in fig.~\ref{fig:MCHMnow}. 
Note that these lines do not exceed $c_{i}=1$ because of the positivity of $\xi$. 
A simple estimate of the allowed size of $\xi$ can therefore be obtained from the intersection of these $\xi$-lines with the corresponding contours in the $c_i$ parameter space. From the intersection with the 95\% probability contours of current Higgs limits, we find that the parameter $\xi$ cannot exceed 0.22 (0.12) in the model CHM-4 (CHM-5). This bound stems from the Higgs coupling to $t$ and $b$ ($t$) and can be translated into a minimal new-physics scale $f$ of $530$ ($710$) GeV, in agreement with \cite{Sanz:2017tco}. 

Moving onto the expected sensitivities at future colliders, we can use the projected experimental limits on the parameters in fig.~\ref{fig:Future} to quantify the attainable impact on composite Higgs scenarios.~\footnote{See also~\cite{Gu:2017ckc} for a different study of the reach of future Higgs factories, based on slightly different assumptions.}
In the top row of fig.~\ref{fig:MCHMfuture} we compare the current limits from fig.~\ref{fig:MCHMnow} focussing on the projected HL-LHC limits in the $c_\psi$ vs.~$c_V$ planes for third generation $\psi$.
We find future HL-LHC bounds of $\xi<0.10$ and $\xi<0.042$ for the CHM-4 and CHM-5 scenarios, respectively.
These limits stem from the 95.4\% probability boundaries for $c_t$ and $c_b$ and can be translated into $f>770$ GeV and $f>1200$ GeV.
The second row shows the magnification of the grey frames from the first row, to focus on the expected precisions at future lepton colliders. 
The resulting estimates on the size of $\xi$ and $f$ are given in table~\ref{tab:CHMresults}, where we also show for comparison the current and future estimates from the LHC. As in table~\ref{tab:futureresults}, the results for each future lepton collider are shown in combination with the HL-LHC, and their individual limits in parentheses.
From that table we observe that all future lepton machines would be able to test values of $\xi$ up to $O(10^{-2})$ or, equivalently, $f$ scales of the order to 3 to 4 TeV.
\begin{table}[t!]
\begin{center}
\begin{tabular}{ c c | c | c c c c c c c}
\toprule
Model& Collider 
& LHC & \cellcolor{color1}HL-LHC & \cellcolor{color2}ILC & \cellcolor{color2}ILC & \cellcolor{color3}CLIC & \cellcolor{color3}CLIC & \cellcolor{color4}CEPC & \cellcolor{color5}FCC-ee \\[-2pt]
& & now & \cellcolor{color1} & \cellcolor{color2}250 & \cellcolor{color2}all & \cellcolor{color3}380 & \cellcolor{color3}all & \cellcolor{color4} & \cellcolor{color5} \\[-2pt]
&$L$ [ab$^{-1}$] & 0.06 & \cellcolor{color1}3 & \cellcolor{color2}1.2 & \cellcolor{color2}5.3 & \cellcolor{color3}0.5 & \cellcolor{color3}4 & \cellcolor{color4}5 & \cellcolor{color5}12.6 \\
\midrule
CHM-4 & $\xi~\![\times 10^{-3}]$ & 220 & 100 & 12 & 5.1 & 18 & 7.5 & 6.0 & 4.3 \\
& & & & (13) & (5.4) & (21) & (9.2) & (6.2) & (4.8) \\
& $f$ [GeV] & 530 & 770 & 2300 & 3400 & 1800 & 2800 & 3200 & 3800 \\
& & & & (2200) & (3300) & (1700) & (2500) & (3300) & (3500) \\
\midrule
CHM-5 & $\xi~\![\times 10^{-3}]$ &120 & 42 & 8.0 & 3.5 & 12 & 4.6 & 4.3 & 3.1 \\
& & & & (8.9) & (3.6) & (15) & (5.2) & (4.7) & (3.2) \\
& $f$ [GeV] & 710 & 1200 & 2800 & 4200 & 2300 & 3600 & 3800 & 4400 \\
& & & & (2600) & (4100) & (2000) & (3400) & (3600) & (4300) \\
\bottomrule
\end{tabular}
\caption{Estimates on the size the CHM ratio $\xi=v^2/f^2$ from the intersection of the $\xi$-lines with the 95.4\% probability $c_i$-contours. We also translate the result into the corresponding value of the symmetry breaking scale $f$. The ILC, CLIC, CEPC and FCC numbers also include the final HL-LHC results; their individual limits are given in parentheses.
}
\label{tab:CHMresults}
\end{center}
\end{table}

          
\section{Conclusions}
\label{sec:conclusions}

The discovery of a 125 GeV mass scalar at the LHC immediately raised the question of whether the newly-discovered particle was the SM Higgs.
To clarify this, precise measurements of the properties of the scalar particle are needed, in the same way that the precision tests of the properties of the $Z$ boson were crucial in the determination of the validity of the SM description of the electroweak interactions.
In order to fully determine the nature of the Higgs-like particle, i.e.~whether it is a doublet or not, one needs, in particular, to measure
the correlations between single-Higgs and multi-Higgs processes. Currently, however, only accurate information about the former can be experimentally accessed. This is still enough to at least address the question of whether new physics may be hiding
in the basic single-Higgs couplings. In this paper we have addressed this particular question, using the general formalism of the
Higgs electroweak chiral Lagrangian, and updated the current knowledge about single-Higgs interactions using a fit including
the latest LHC results from run 2.

We have performed a global fit of the electroweak chiral Lagrangian Wilson coefficients, $c_i$, to the currently available Higgs signal strength measurements. 
First, we have discussed the overall constraining power of the data, explaining the importance of using priors to ``regulate'' the coefficients that are currently weakly bounded and that can compensate the effect of other interactions, leading to an overfitting problem. The charm Wilson coefficient is one of these couplings. Not only the direct bounds on $h\to \bar{c}c$ are poor but also $c_c$ can have a sizable impact in the Higgs width and important loop processes, e.g.~gluon fusion. As we explained, a flat prior allowing large values, of $O(1-10)$, for this parameter would ``artificially'' pull all the \ewXL~couplings toward values larger than the SM without conflict with Higgs observables. We therefore set a Gaussian prior for $c_c$ to contain it within the natural EFT region. 
Still, our results show that regions of the parameter space far away from the SM are allowed, coming from accidental symmetries in the signal strength formulas. These regions are identified to be around $-1$ for the fermion and vector boson coupling parameters, around $c_g\approx \pm1.5$ and the photon coupling can also have values of roughly $\pm 1.5$, $\pm 7$ or $\pm 8.5$. Such deviations are, however, not expected to be consistent with the deviations predicted by the EFT approach.

In order to study more in detail the EFT natural region, we isolate it by using minimal priors for all the parameters.
Apart from $c_c$, we also apply a Gaussian prior on $c_{Z\gamma}$. (In this case, the use of a flat prior is less problematic than for $c_c$ and would result in an upper bound $|c_{Z\gamma}|\lesssim 35$.) All the other parameters have a flat prior in the range $\pm 1$ around their SM values in order to cut away all solutions not consistent with the EFT power counting.
The results of this fit are summarized in table~\ref{tab:fitresults}. They illustrate to what extent the data is consistent with the SM hypothesis, and what the allowed size of new physics effects in the Higgs couplings is. Indeed, the SM limit of all individual $c_i$ is mostly compatible with the results of the fit at the 68\% probability level. With this statistical significance, the vector boson coupling has an uncertainty of 6\% and the third generation fermion and the gluon couplings can deviate at most by 7\% to 13\% from their SM limits. The muon coupling features the largest discrepancy of all parameters between the SM and the fit result, but its uncertainty of 40\% is still sizeable. Also the photon coupling has a rather large uncertainty of 20\%.
Again, we note that these results only have implications regarding how large new physics in single-Higgs couplings can be, but this level of consistency between the data and the SM has no direct implications from the point of view of whether the Higgs is a singlet or a doublet of $SU(2)_L$.

After studying in detail the sensitivity to new physics of current data, we have used the future projections of the high-luminosity upgrade of the LHC to quantify how precisely the $c_i$ will be determined at the end of the LHC era. Applying flat priors around the SM values for all parameters in eq.~\eqref{eq:2}, we find the following HL-LHC sensitivities:
the vector boson coupling could be tested with a precision of 3\%, the third generation fermion couplings as well as the gluon and muon couplings will be known up to roughly 5\%. The sensitivity to a direct Higgs coupling to two photons would be at the level of 7.5\%, while the coupling to one photon and one $Z$ boson could be extracted with an uncertainty of 95\%.

Apart from these bounds, we have also discussed the potential sensitivity to the \ewXL~Wilson coefficients at future lepton colliders. 
Linear collider concepts like ILC and CLIC are each able to measure $c_V$ at the sub-percent level and $c_b$, $c_c$ and $c_\tau$ at the percent level during their low-energy runs. 
Summing up all further data from potential runs at higher centre-of-mass energies, these bounds decrease to few per mil for $c_V$ and $c_{b}$ and to the percent level for $c_c$ and $c_\tau$. A precision of a few percent would be attainable on $c_t$ and $c_g$, and of order 10\% on $c_\gamma$ and $c_\mu$.
We contrast this with the CEPC and FCC-ee projections based on circular accelerators. The latter could also reduce the uncertaintiy of $c_c$ below 1\%, and to the few per mil for $c_V$, $c_b$ and $c_\tau$. The precision on $c_\mu$ would be similar to the one obtained at the HL-LHC.
Taking into account the differences in the integrated luminosities, one can see that all future concepts would be able to measure the Higgs coupling to vector bosons, bottoms, charms, taus with similar precision. 
This is not the case, however, if linear colliders only run at the low-energy stage, in which case they cannot compete in precision with the circular colliders sensitivities. 
On the other hand, the high-energy runs would also allow the linear options to constrain $c_t$, $c_g$ and $c_\gamma$ separately to a good accuracy. The absence of a direct handle to $tth$ production at the energies the circular colliders will operate, however, restricts them only to constrain certain linear combinations of these parameters. 

Finally, we further analysed the implications of our model-independent results for two manifestations of the minimal composite Higgs models.
Their characteristic parameter $\xi$ is found to be smaller than $0.22$ or $0.12$, depending on whether the fermions are embedded in the representations $\mathbf{4}$ or $\mathbf{5}$, respectively. This translates into lower bounds of the typical symmetry breaking scale $f$ around $530$ GeV and $710$ GeV. Such limits could be further extended to the multi-TeV range at future lepton colliders.


\acknowledgments
We thank A. Pich for fruitful discussions. We further gratefully acknowledge the discussion of preliminary results at the Higgs Couplings 2017 conference. We thank the INFN Roma Tre Cluster, where the fits were performed.
The work of OE and CK was supported by the Spanish Government and ERDF funds from the European Commission (Grants No. FPA2014-53631-C2-1-P and SEV-2014-0398). CK acknowledges the support of the Alexander von Humboldt Foundation. CK thanks Fermilab and the Enrico Fermi Institute at the University of Chicago, and OE the INFN in Rome for hospitality, where parts of this research were carried out. This manuscript has been authored by Fermi Research Alliance, LLC under Contract No. DE-AC02-07CH11359 with the U.S. Department of Energy, Office of Science, Office of High Energy Physics.

\appendix

\newpage

\section{Theoretical expressions for Higgs observables}
\label{app:ThExpr}

In this appendix we list all the relevant formulae for the calculation of the Higgs signal strengths, $\mu_{X\rightarrow h\rightarrow Y}$, for the different production modes $X \in \{$ggF, Vh, VBF, tth$\}$ and decay channels $Y \in \{WW, ZZ, \gamma \gamma, Z\gamma, \bar{b}b, \tau^+\tau^-, \mu^+\mu^- \}$. 
We assume the narrow-width approximation holds for all the values of the \ewXL~Wilson coefficients we consider, and decompose the corresponding cross sections as $\sigma_X \times \text{Br}_Y$.\footnote{See \cite{Campbell:2017rke} for an example of finite-width effects on the measurements of Higgs on-shell rates.} 

Using the Lagrangian in eq.~\eqref{eq:2}, we find, at the leading-order,
\begin{align}
\frac{\sigma(\text{ggF})}{\sigma(\text{ggF})_{\text{SM}}}\simeq    \frac{\Gamma^{gg}}{\Gamma^{gg}_{\text{SM}}}, \quad
\frac{\sigma(\text{VBF})}{\sigma(\text{VBF})_{\text{SM}}} = \frac{\sigma(\text{Vh})}{\sigma(\text{Vh})_{\text{SM}}} = c_V^2, \quad
\frac{\sigma(\text{tth})}{\sigma(\text{tth})_{\text{SM}}} =  c_{t}^2 ,
\end{align}
where we use $\Gamma^Y$ to denote the partial width of the $h$ decay to $Y$. 

The tree-level decay rates of $h$ decays to massive gauge bosons $V=W,Z$ and to fermions $\psi=b,c,\tau,\mu$ get rescaled compared to the SM by a factor of $c_V^2$ and $c_{\psi}^2$, respectively,
\begin{align}
\frac{\Gamma^{ZZ}}{\Gamma^{ZZ}_{\text{SM}}} = \frac{\Gamma^{WW}}{\Gamma^{WW}_{\text{SM}}} =  c_{V}^2, \quad
\frac{\Gamma^\psi}{\Gamma^\psi_{\text{SM}}} =  c_{\psi}^2.
\end{align}
For the decays into $\gamma\gamma$, $gg$ and $Z\gamma$ we use the loop expressions \cite{Gunion:1989we,Manohar:2006gz,Contino:2014aaa},
\begin{align}
  \frac{\Gamma^{\gamma \gamma}}{\Gamma^{\gamma \gamma}_{\text{SM}}} &=    \frac{  \left|  \sum\limits_{q}   \frac{4}{3} N_C  Q_q^2    c_{q} A_{1/2}(x_q) \eta^{q,\gamma\gamma}_{\text{QCD}}   + \sum\limits_{f=\tau,\mu,e}\frac{4}{3} c_{f}  A_{1/2}(x_{f})    + c_V A_{1}(x_W) + 2 c_{\gamma}  \right|^2    }{ \left| \sum\limits_{q}   \frac{4}{3} N_C  Q_q^2 A_{1/2}(x_q) \eta^{q,\gamma\gamma}_{\text{QCD}}   + \frac{4}{3}  A_{1/2}(x_{\tau})    +  A_{1}(x_W)   \right|^2    }   \,, \\
   \frac{\Gamma^{gg}}{\Gamma^{gg}_{\text{SM}}}  &=  \frac{   \left|  \sum\limits_{q}  \frac{1}{3} c_q A_{1/2}(x_q) \eta^{q,gg}_{\text{QCD}}   + \frac{1}{2}  c_{g}      \right|^2     }{   \left| \sum\limits_{q}  \frac{1}{3}  A_{1/2}(x_q) \eta^{q,gg}_{\text{QCD}}    \right|^2    }   \,, \\
 \frac{\Gamma^{Z \gamma}}{\Gamma^{Z \gamma}_{\text{SM}}} &= \frac{\left| \sum\limits_{\psi} c_{\psi} N_C  Q_{\psi}A_{\psi}(x_{\psi},\lambda_{\psi}) \eta^{\psi,Z\gamma}_{\text{QCD}}+ c_{V}A_{W}(x_{W},\lambda_{W}) +c_{Z\gamma}\right|^{2}}{\left| \sum\limits_{\psi}  N_C  Q_{\psi}A_{\psi}(x_{\psi},\lambda_{\psi})+ A_{W}(x_{W},\lambda_{W}) \right|^{2}},~~~~~~~~~~~~~~~~~~
\end{align}
which also include the tree-level contributions from $c_\gamma$, $c_g$ and $c_{Z\gamma}$.
In the equations above, $x_i =  4 m_i^2/m_h^2$, $\lambda_{i} = 4 m_i^2/m_Z^2$, and $Q_{\psi}$ is the electric charge of a fermion $\psi$. The $\eta^{x,Y}_{\text{QCD}}$ are QCD corrections of $\mathcal{O}(\alpha_s)$. We only take into account ${\eta^{t,gg}_{\text{QCD}}  = 1+11\alpha_{s}/4\pi}$ and  $\eta^{t,\gamma\gamma}_{\text{QCD}} =\eta^{t,Z\gamma}_{\text{QCD}} = 1 -\alpha_{s}/\pi$. Other contributions~\cite{Manohar:2006gz,Contino:2013kra,Contino:2014aaa} have a very small effect in our results and we neglect them. 
The one-loop functions are 
\begin{align}
    A_{1/2}(x) &=   \frac{3}{2}  x \left[    1 + (1-x) f(x) \right] \, , \nonumber \\
    A_{1}(x)  &=  - [  2 + 3 x + 3 x (2-x) f(x)   ] \, , \nonumber \\
    A_{\psi}(x_{\psi},\lambda_{\psi}) & = -2 \frac{T^{3}_{\psi}-2Q_{\psi}\sin^{2}{\theta_{w}}}{\sin{\theta_{w}}\cos{\theta_{w}}} \left[I_1\left(x_{\psi},\lambda_{\psi}\right)-I_2\left(x_{\psi},\lambda_{\psi}\right)\right]\, ,\nonumber \\
    A_{W}(x_{W},\lambda_{W}) & = - \cot{\theta_{w}}\Big[4(3-\tan^{2}{\theta_{w}})I_{2}(x_{W},\lambda_{W}) \nonumber \\
&\qquad+\left((1+\tfrac{2}{x_{W}})\tan^{2}{\theta_{w}}-(5+\tfrac{2}{x_{W}})\right) I_{1}(x_{W},\lambda_{W})\Big] \, ,
\end{align}
with $T^{3}_{\psi}$ being the third component of the weak isospin of the fermion $\psi$, $\theta_{w}$ the weak mixing angle, and
\begin{align}
I_1\left(\tau,\lambda\right)&=\frac{\tau\lambda}{2\left(\tau-\lambda\right)}+\frac{\tau^2\lambda^2}{2\left(\tau-\lambda\right)^2}\left[f\left(\tau\right)-f\left(\lambda\right)\right]+\frac{\tau^2\lambda}{\left(\tau-\lambda\right)^2}\left[g\left(\tau\right)-g\left(\lambda\right)\right], \nonumber \\
I_2\left(\tau,\lambda\right)&=-\frac{\tau\lambda}{2\left(\tau-\lambda\right)}\left[f\left(\tau\right)-f\left(\lambda\right)\right] \; .
\end{align}
Finally, the functions $f(x)$ and $g(x)$ read
\begin{align}
g\left(x\right)\; &=\; \begin{cases}
\sqrt{x-1}\, \arcsin (1/\sqrt{x}) & x\geq 1 \\
\frac{\sqrt{1-x}}{2}\left[\ln\frac{1+\sqrt{1-x}}{1-\sqrt{1-x}}-i\pi\right] \,\quad& x<1 \, \end{cases}, \nonumber \\
f(x)\; &=\; \begin{cases} \arcsin^2(1/\sqrt{x})\,  \quad & x\geq 1 \\[3pt]   - \dfrac{1}{4} \Big[\ln\Big( \frac{1+\sqrt{1-x}}{1-\sqrt{1-x}}\Big)- i\pi \Big]^2\,  & x<1 \end{cases}. \, 
\end{align}


\newpage

\section{Relation to the $\kappa$-framework}
\label{app:kappa}

The so-called $\kappa$-framework was introduced as a recommendation from the LHC Higgs cross section working group, to explore deviations of the couplings of a Higgs-like particle with respect to the SM~\cite{LHCHiggsCrossSectionWorkingGroup:2012nn,Heinemeyer:2013tqa}. While one of its goals is to avoid reference to specific models, one significant (simplifying) assumption is that the tensor structure of the Higgs couplings, and therefore the kinematic distributions of Higgs processes, are SM like. In other words, in this framework only modifications of the SM coupling strengths are considered. These are parameterized via scale factors, denoted as $\kappa_i$. Moreover, these coupling modifiers are defined ``phenomenologically'', in the sense that each $\kappa_i$ is defined as ratios of cross sections and decay widths:
\begin{equation}
  \label{eq:kappa1}
  \kappa^{2}_{X} = \frac{\sigma(X\rightarrow h)}{\sigma(X\rightarrow h)_{\text{SM}}}, \qquad \kappa^{2}_{Y} = \frac{\Gamma(h\rightarrow Y)}{\Gamma(h\rightarrow Y)_{\text{SM}}},
\end{equation}
so the SM is recovered for $\kappa_i=1$. While eq.~(\ref{eq:kappa1}) may resemble some of the \ewXL~corrections described in the previous appendix, the 2 formalisms are fundamentally different. 
Indeed, the Wilson coefficients of the \ewXL~are introduced at the Lagrangian level, in a well-defined theory where one can compute predictions at any order in perturbation theory. In the $\kappa$-formalism, on the other hand, higher-order accuracy is lost for $\kappa_i\not =1$. In this appendix, we briefly comment on the connection between the two approaches. 

From the point of view of how both approaches describe new physics effects in the data, the main practical difference between the \ewXL~and the $\kappa$-framework is in the treatment of the one-loop induced processes. 
The $\kappa$-formalism allows to express the couplings associated to loop-induced processes as a function of the $\kappa_i$ couplings of the particles running in the loops. However, in the general effective treatment which allows, e.g.~new particle effects in the loops, $\kappa_{\gamma}$ and $\kappa_g$ are treated as independent parameters in the fits. This is the scenario we consider here. In the \ewXL~approach, on the other hand, contributions from modified couplings on the loops and new local corrections are parameterized separately. From this point of view, the \ewXL~ is clearly a better way of parameterizing new physics effects in the data, as it provides a cleaner separation of the origin of new physics effects with the same number of parameters~\cite{Buchalla:2015wfa}.

The mapping from the Wilson coefficients $c_{i}$ to the $\kappa_{i}$ parameters is well defined using the relations of appendix~\ref{app:ThExpr}. These relations can be written as
\begin{equation}
\label{eq:kappa2}
  \kappa_{i} =  |f_i(c_{j})| \equiv \frac{|{\cal A}_i(c_{j})|}{|{\cal A}_i(c_{j}^\SM)|}, 
\end{equation}
where ${\cal A}$ is the corresponding transition amplitude of each process. 
The absolute value on the right hand side is necessary, as the loop functions of the light fermions ($b,\tau,\mu,\dots$) for the $\kappa_{\gamma}$ and $\kappa_{g}$ are complex. 

The inverse of eq.~\eqref{eq:kappa2} is, however, not a well-defined function. We can still obtain an approximate inverse, to connect both formalisms in the opposite direction. 
This can be easily obtained if we assume that all the imaginary parts are negligible. 
While this is a good approximation for some of the coefficients in $f_i(c_{j})$, for example for the coefficient of $c_{t}$, it is not the case for the coefficients of the light fermion loops, where real and imaginary parts are of similar size. Nevertheless, as long as the Wilson coefficients stay relatively close to the SM value, neglecting the imaginary parts completely is still a good approximation, because in $\kappa_g$ ($\kappa_\gamma$) the real part of the top loop (top and $W$ loops) contribution dominates over all the other terms.

With the assumption of vanishing imaginary parts, eq.~\eqref{eq:kappa2} becomes
\begin{equation}
  \label{eq:kappatrans}
  \begin{pmatrix}
    \kappa_{V}\\
    \kappa_{t}\\
    \kappa_{b}\\
    \kappa_{\ell}\\
    \kappa_{g}\\
    \kappa_{\gamma}
  \end{pmatrix}
  = 
  \begin{pmatrix}
    1 & 0 & 0 & 0 & 0 & 0 \\
    0 & 1 & 0 & 0 & 0 & 0 \\
    0 & 0 & 1 & 0 & 0 & 0 \\
    0 & 0 & 0 & 1 & 0 & 0 \\
    0 & 1.055 & -0.055 & 0 & 1.3891 & 0 \\
    1.2611 & -0.2683 & 0.0036 & 0.0036 & 0 & -0.3039 \\
  \end{pmatrix}
  \cdot
  \begin{pmatrix}
    c_{V}\\
    c_{t}\\
    c_{b}\\
    c_{\tau}\\
    c_{g}\\
    c_{\gamma}
  \end{pmatrix}.
\end{equation}
We checked the validity of the approximation of vanishing imaginary parts by translating the central values of $c_i$ coefficients from our fit into the $\kappa_i$ parameters. We used both eq.~\eqref{eq:kappatrans} and the exact expressions that can derived from the equations in appendix~\ref{app:ThExpr}. As expected, the dominance of the top and $W$ loops results in both approaches giving the same result with negligible differences.

\begin{figure}[t]
   \begin{picture}(400,400)(0,0)
      \put(0,0){\includegraphics[width=438pt]{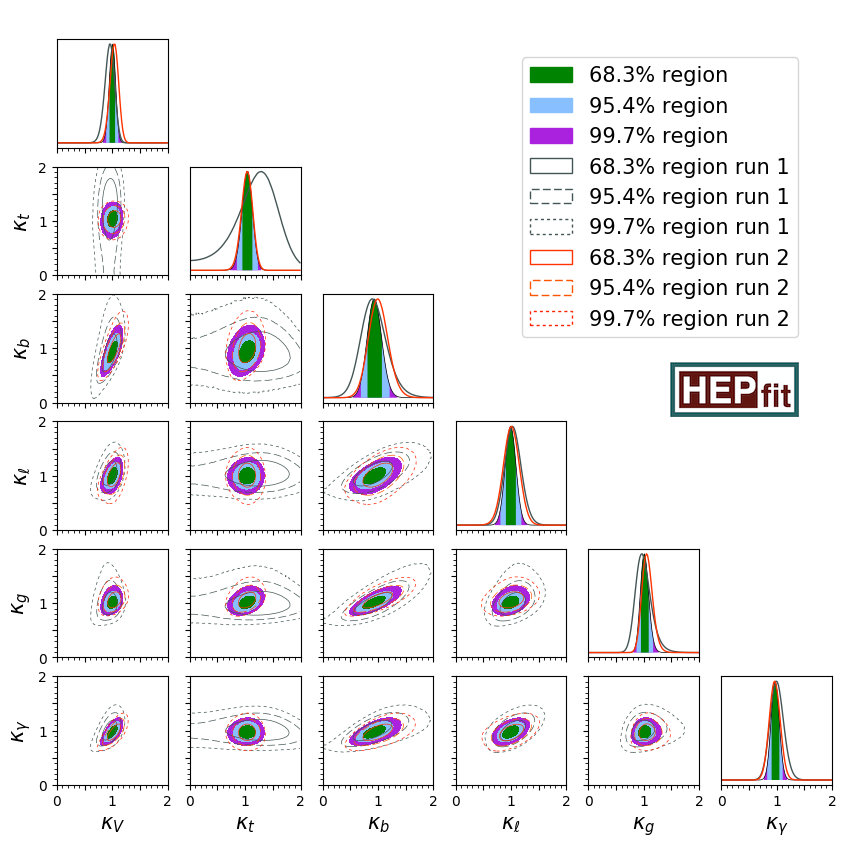}}
   \end{picture}
   \caption{For the parameters $\kappa_{i}$ with $i = V, t, b, \ell, g, \gamma$ we display the one-dimensional posterior distribution as well as their two-dimensional correlations. The regions allowed at $68.3\%$, $95.4\%$ and $99.7\%$ probability by current Higgs data are represented by the green, blue and purple filled contours, respectively. Additionally, we show the single contributions from pre-13 TeV run data (dark gray) and LHC run 2 data (orange).}
   \label{fig:Kappas}
\end{figure}

The inverse of eq.~\eqref{eq:kappatrans} is
\begin{equation}
  \label{eq:kappatrans.inv}
  \begin{pmatrix}
    c_{V}\\
    c_{t}\\
    c_{b}\\
    c_{\tau}\\
    c_{g}\\
    c_{\gamma}
  \end{pmatrix}
  = 
  \begin{pmatrix}
    1 & 0 & 0 & 0 & 0 & 0 \\
    0 & 1 & 0 & 0 & 0 & 0 \\
    0 & 0 & 1 & 0 & 0 & 0 \\
    0 & 0 & 0 & 1 & 0 & 0 \\
    0 & -0.76 & 0.04 & 0 & 0.72 & 0 \\
    4.15 & -0.88 & 0.012 & 0.012 & 0 & -3.29 \\
  \end{pmatrix}
  \cdot
  \begin{pmatrix}
    \kappa_{V}\\
    \kappa_{t}\\
    \kappa_{b}\\
    \kappa_{\ell}\\
    \kappa_{g}\\
    \kappa_{\gamma}
  \end{pmatrix}.
\end{equation}
With these relations one can translate the results of a $\kappa_i$ fit into the \ewXL~formalism and vice-versa. 
In order to do so, however, it is important to have all the relevant information about the fits. In particular,
the median and errors of the parameters are not sufficient, since there may be also significant correlations between them.  
For instance, the results of the $\kappa$-fit to the data of table~\ref{tab:signal_strengths} are shown in table~\ref{tab:kappa.c} 
and figure~\ref{fig:Kappas}. From the figure, one can see the existence of significant correlations between, e.g., $\kappa_V$ and $\kappa_\gamma$.
Ignoring this and using eq.~(\ref{eq:kappatrans.inv}) to translate the $\kappa_i$ results into the $c_i$ parameterization
would result in a significant increase on the uncertainty on $c_\gamma$, compared to fitting directly
to the \ewXL. This can be understood from the large $(6,1)$ matrix element in eq.~(\ref{eq:kappatrans.inv}),
and is illustrated in the last two columns in table~\ref{tab:kappa.c}. These columns show 
the results of the direct $c_i$ fit and the translation of the $\kappa_i$ one using eq.~(\ref{eq:kappatrans.inv}) and ignoring
correlations. Using the full correlation matrix of the $\kappa_i$ fit,
\begin{equation}
\rho_{\kappa}=
\left(\begin{array}{c c c c c c}
        1~~&~~ 0.13~~&~~ 0.73~~&~~ 0.49~~&~~ 0.26~~&~~ 0.68\\
        0.13~~&~~ 1~~&~~ 0.27~~&~~ 0.06~~&~~ 0.33~~&~~ 0.05\\
        0.73~~&~~ 0.27~~&~~ 1~~&~~ 0.56~~&~~ 0.74~~&~~ 0.60\\
        0.49~~&~~ 0.06~~&~~ 0.56~~&~~ 1~~&~~ 0.33~~&~~ 0.47\\ 
        0.26~~&~~ 0.33~~&~~ 0.74~~&~~ 0.33~~&~~ 1~~&~~ 0.11\\ 
        0.68~~&~~ 0.05~~&~~ 0.60~~&~~ 0.47~~&~~ 0.11~~&~~ 1
\end{array}
\right),
\end{equation}
one can reproduce the exact $c_i$ results to a good accuracy, with differences given only by the deviations from Gaussianity of the fits. 
The same considerations about the importance of providing all the necessary information to reconstruct the posterior of a fit (at least at the Gaussian level)
applies if one wants to translate the $\kappa_i$ or $c_i$ results in terms of specific models.

\begin{table}[t!]
\begin{center}
\begin{tabular}{ c | c || c | c | c }
\toprule
Parameter & Fit result & Parameter & Fit result & Result from $\kappa$-fit and eq.~\eqref{eq:kappatrans.inv}\\
                  &                &                  &                & (ignoring $\kappa$ correlations)\\
  \midrule
  $\kappa_V$ & $1.00\pm 0.06$ & $c_V$ & $1.00\pm 0.06$ & $1.00\pm 0.06$ \\
$\kappa_t$ & $1.04^{+0.09}_{-0.10}$ & $c_t$ & $1.03\pm 0.09$ & $1.04\pm 0.10$\\
$\kappa_b$ & $0.94\pm 0.13$ & $c_b$ & $0.94\pm 0.13$ & $0.94\pm 0.13$ \\
$\kappa_\ell$ & $1.00\pm 0.10$ & $c_\tau$  & $1.01\pm 0.10$ & $1.00\pm 0.10$ \\
$\kappa_g$ & $1.02^{+0.08}_{-0.07}$ & $c_g$ & $-0.01^{+0.08}_{-0.07}$& $-0.02\pm 0.10$\\
$\kappa_\gamma$ & $0.97\pm 0.07$ & $c_\gamma$ &  $0.05\pm 0.20$ & $0.06\pm 0.35$ \\
\bottomrule
\end{tabular}
\caption{For the $\kappa$ formalism, we list the values of the median and the limits of its 68\% probability interval. They are confronted with the respective values for a fit to the parameters of the chiral Lagrangian~\eqref{eq:2} with $c_{c}=c_{\mu}=1$ and $c_{Z\gamma}=0$. In the last column we show the values of the $c_{i}$ when the result of the $\kappa_{i}$ fit is translated using eq.~\eqref{eq:kappatrans.inv} and ignoring the correlations in the output of the $\kappa_i$ fit. The difference in the error obtained for $c_\gamma$ illustrates the importance of such correlations in the fit.}
\label{tab:kappa.c}
\end{center}
\end{table}


\clearpage

\section{Projected uncertainties at future colliders}
\label{app:futcol_data}

In this appendix we collect the different inputs used in the analysis presented in Section~\ref{sec:projections}. For the HL-LHC projections, we report in table~\ref{tab:HLLHCinputs} the uncertainties on the Higgs signal strength that could be measured using in the different categories defined in the ATLAS Refs.~\cite{ATL-PHYS-PUB-2013-014,ATL-PHYS-PUB-2014-011,ATL-PHYS-PUB-2014-016}. From CMS we use the signal strengths per decay mode given in Ref.~\cite{CMS-NOTE-2013-002}. In both cases we use the numbers corresponding to the most optimistic scenario in terms of theoretical uncertainties. For the projections at future lepton colliders in Table~\ref{tab:futureinputs}, the projected uncertainties are separated according to the main production mechanism: associated production with a $Z$ boson ($Zh$), $e^+ e^- \to \nu \bar{\nu} h$ via $W$ boson fusion (WBF), or associated production with a $t\bar{t}$ pair ($tth$).

\begin{table}[h]
\begin{center}
\begin{tabular}{c c}
{\small
\begin{tabular}{ c c | c}
\toprule
&&  {\cellcolor{color1}HL-LHC (ATLAS)} \\[-1pt]
&&  {\cellcolor{color1} 3000 ab$^{-1}$} \\[-1pt]
Decay&Category &  Uncertainty [$\times 10^{-3}$]\\
\midrule
& 0j &   50\\
$WW$& 1j &  110\\
& 2j &   90\\
\midrule
 & ggF-like &   50\\
 & VBF-like &   160\\
$ZZ$& Wh-like &   170\\
 & Zh-like &  170\\
 & tth-like &   170\\
\midrule
$\tau\tau$& VBF-like &  150\\
\midrule
$\mu\mu$& incl. &   140\\
& tth-like &  230\\
\midrule
$bb$& WH-like &   360\\
 & ZH-like &   130\\
\midrule
 & 0j &   50\\
 & 1j &   60\\
$\gamma\gamma$& VBF-like &   150\\
 & Wh-like &  180\\
 & Zh-like &   280\\
 & tth-like &   120\\
\midrule
$Z\gamma$& incl. & 270\\
\bottomrule
\end{tabular}
}
&
{\small
\begin{tabular}{ c | c}
\toprule
& {\cellcolor{color1}HL-LHC (CMS)} \\[-1pt]
& {\cellcolor{color1} 3000 ab$^{-1}$} \\[-1pt]
Decay&   Uncertainty [$\times 10^{-3}$]\\
\midrule
$WW$& 40\\
\midrule
$ZZ$& 40\\
\midrule
$\tau\tau$& 50\\
\midrule
$\mu\mu$&  200\\
\midrule
$bb$& 50\\
\midrule
$\gamma\gamma$& 40\\
\midrule
$Z\gamma$& 200\\
\bottomrule
\end{tabular}
}
\end{tabular}
\caption{Inputs for the Higgs-boson signal strength's uncertainties at the HL-LHC from ATLAS~\cite{ATL-PHYS-PUB-2013-014,ATL-PHYS-PUB-2014-011,ATL-PHYS-PUB-2014-016} and CMS~\cite{CMS-NOTE-2013-002}.}
\label{tab:HLLHCinputs}
\end{center}
\end{table}

\begin{table}[t!]
\begin{center}
{\small
\begin{tabular}{ c c | c c c c c c c c c}
\toprule
& Collider
& \multicolumn{3}{c}{\cellcolor{color2} ILC} & \multicolumn{3}{c}{\cellcolor{color3} CLIC} & \cellcolor{color4}CEPC & \multicolumn{2}{c}{\cellcolor{color5} FCC-ee} \\[-2pt]
& $\sqrt{s}$ [GeV]& \cellcolor{color2}250 & \cellcolor{color2}500 & \cellcolor{color2}1000 & \cellcolor{color3}380 & \cellcolor{color3}1400 & \cellcolor{color3}3000 & \cellcolor{color4} & \cellcolor{color5} 240 & \cellcolor{color5} 350 \\[-2pt]
& $L$[ab$^{-1}$]& \cellcolor{color2}1.2 & \cellcolor{color2} 1.6 & \cellcolor{color2} 2.5 & \cellcolor{color3}0.5 & \cellcolor{color3} 1.5 & \cellcolor{color3} 2 & \cellcolor{color4}5 & \cellcolor{color5} 10 & \cellcolor{color5} 2.6 \\
Decay & Production&  &    &    &  &   &   &   &    &    \\
\midrule
\multirow{2}{*}{$WW$} &$Zh$ &  30 & 51 & & 51 & & & 15 & 9 & \\
 &WBF & & 13 & 10 & & 10 & 7 & & & \\
\midrule
\multirow{2}{*}{$ZZ$} &$Zh$ &  88 & 140 & & & & & 43 & 31 & \\
&WBF & & 46 & 26 & & 56 & 37 & & & \\
\midrule
\multirow{3}{*}{$\bar{b}b$} &$Zh$ &  5.6 & 10 & & 8.6 & & & 2.8 & 2 & \\
&WBF & 49 & 3.7 & 3 & 19 & 4 & 3 & 28 & 22 & 6\\
&$tth$ & & 160 & 38 & & 80 & & & & \\
\midrule
\multirow{2}{*}{$\bar{c}c$} &$Zh$ &  39 & 72 & & 140 & & & 22 & 12 &\\
 &WBF & & 35 & 20 & 260 & 61 & 69 & & & \\
\midrule
\multirow{2}{*}{$\tau^+\tau^-$} &$Zh$ &  20 & 30 & & 62 & & & 12 & 7 & \\
 &WBF & & 50 & 20 & & 42 & 44 & & & \\
\midrule
 \multirow{2}{*}{$\mu^+\mu^-$} &$Zh$ & & & & & & & 170 & 130 & \\
 &WBF & & & 200 & & 380 & 250 & & & \\
\midrule
 \multirow{2}{*}{$gg$} &$Zh$ & 33 & 60 & & 61 & & & 16 & 14 & \\
 &WBF & & 23 & 14 & 100 & 50 & 43 & & & \\
\midrule
\multirow{2}{*}{$\gamma\gamma$} & $Zh$ &  160 & 190 & & & & & 90 & 30 & \\
 &WBF & & 130 & 54 & & 150 & 100 & & & \\
\midrule
$Z\gamma$ & WBF &  & & & & 420 & 280 & & & \\
\midrule
Inclusive & Zh &12  &17 & &16.5 &   &   & 5.1& 4& \\
\midrule
$ $ & &\multicolumn{9}{c}{Uncertainties [$\times 10^{-3}$]}\\
\bottomrule
\end{tabular}
}
\caption{Inputs for the Higgs signal strength's uncertainties at ILC~\cite{Dawson:2013bba}, CLIC~\cite{Dawson:2013bba,Abramowicz:2016zbo}, CEPC~\cite{CEPC-SPPCStudyGroup:2015csa} and FCC-ee~\cite{Gomez-Ceballos:2013zzn}.}
\label{tab:futureinputs}
\end{center}
\end{table}


\clearpage

\bibliography{ewXLHEPfit2017}
\bibliographystyle{JHEP}

\end{document}